\documentclass[11pt,twoside,journal,onecolumn,draftcls]{IEEEtran}

\usepackage{times,graphics,amssymb,array,amsmath,bm,cite}
\usepackage{algorithmic}
\usepackage{algorithm}
\usepackage[dvips]{graphicx}
\usepackage[active]{srcltx}  %SRC Specials for DVI search
\usepackage{epsfig,latexsym}
\usepackage{color}

\newcommand{\mat}{\boldsymbol}

\newcommand{\eeq}{\end{equation}}

\newcommand{\bT}{\mbox{\boldmath $T$}}

\newcommand{\bs}{\mbox{\boldmath $s$}}

\newcommand{\bbf}{\mbox{\boldmath $f$}}

\newcommand{\bM}{\mbox{\boldmath $M$}}

\newcommand{\bA}{\mbox{\boldmath $A$}}

\newcommand{\bX}{\mbox{\boldmath $X$}}

\newcommand{\bH}{\mbox{\boldmath $H$}}

\newcommand{\blambda}{\mbox{\boldmath $\lambda$}}
\newcommand{\bZ}{\mbox{\boldmath $Z$}}
\newcommand{\bU}{\mbox{\boldmath $U$}}
\newcommand{\bGamma}{\mbox{\boldmath $\Gamma$}}

\newcommand{\bV}{\mbox{\boldmath $V$}}
\newcommand{\bL}{\mbox{\boldmath $L$}}
\newcommand{\bG}{\mbox{\boldmath $G$}}

\newcommand{\bW}{\mbox{\boldmath $W$}}
\newcommand{\bLambda}{{\bf \Lambda}}

\newcommand{\bell}{\mbox{\boldmath $\ell$}}

\newcommand{\bx}{\mbox{\boldmath $x$}}
\newcommand{\bz}{\mbox{\boldmath $z$}}

\newcommand{\bn}{\mbox{\boldmath $n$}}

\newcommand{\bQ}{\mbox{\boldmath $Q$}}

\newcommand{\bR}{\mbox{\boldmath $R$}}
\newcommand{\bF}{\mbox{\boldmath $F$}}

\newcommand{\bI}{\mbox{\boldmath $I$}}

\newcommand{\by}{\mbox{\boldmath $y$}}
\newcommand{\ds}{\displaystyle}

\newcommand{\bY}{\mbox{\boldmath $Y$}}

\newcommand{\beq}{\begin{equation}}

\newcounter{MYtempeqncnt}
\newtheorem{remark}{Remark}

\newtheorem{lemma}{Lemma}
\newtheorem{proposition}{Proposition}

\def\IR{{\mathbb R}}

\begin{document}

\title{Energy Efficiency Optimization in Relay-Assisted MIMO Systems with Perfect and Statistical CSI}
\author{{Alessio Zappone~\IEEEmembership{Member,~IEEE,} Pan Cao~\IEEEmembership{Student Member,~IEEE,}, and Eduard A. Jorswieck~\IEEEmembership{Senior Member,~IEEE,}}\\
\thanks{Copyright (c) 2013 IEEE. Personal use of this material is permitted. However, permission to use this material for any other purpose must be obtained from the IEEE by sending a request to pubs-permissions@ieee.org.

The authors are with the Technische Universit\"at Dresden, Communications Laboratory, Dresden, Germany, (e-mail: \{Alessio.Zappone,Pan.Cao,Eduard.Jorswieck\}@tu-dresden.de);
Part of this work has been presented at the IEEE China Summit and International Conference on Signal and Information Processing (ChinaSIP 2013). The work of Alessio Zappone has been funded by the German Research Foundation (DFG) project CEMRIN, under grant ZA 747/1-1. The work of Eduard Jorswieck is supported in part by the German Research Foundation (DFG) in the Collaborative Research Center 912 Highly Adaptive Energy-Efficient Computing.}
}\maketitle

\begin{abstract}
A framework for energy-efficient resource allocation in a single-user, amplify-and-forward (AF), relay-assisted, multiple-input-multiple-output (MIMO) system is devised in this paper. Previous results in this area have focused on rate maximization or sum power minimization problems, whereas fewer results are available when bits/Joule energy efficiency (EE) optimization is the goal. Here, the performance metric to optimize is the ratio between the system's achievable rate and the total consumed power. The optimization is carried out with respect to the source and relay precoding matrices, subject to quality-of-service (QoS) and power constraints. Such a challenging non-convex optimization problem is tackled by means of fractional programming and alternating maximization algorithms, for various channel state information (CSI) assumptions at the source and relay. In particular the scenarios of perfect CSI and those of statistical CSI for either the source-relay or the relay-destination channel are addressed.
Moreover, sufficient conditions for beamforming optimality are derived, which is useful in simplifying the system design. Numerical results are provided to corroborate the validity of the theoretical findings.
\end{abstract}
\section{Introduction}\label{Sec:intro}
Wireless relaying is a well-known technique to provide reliable transmission, high throughput, broad coverage and agile frequency reuse in modern wireless networks \cite{Cho2004, Jiang2008}. In a cellular environment, relays are usually deployed in areas where a significant shadowing effect is present such as tunnels or the inside of buildings, as well as in areas that are far away from the transmitter and that otherwise would not be covered. In this context, AF is one of the most widely used choices because it does not require the relays to decode and know the users' codebooks, thus allowing a faster and simpler design and placement of the relays. This relaying strategy is also one candidate approach in the standard LTE-Advanced and is usually referred to as layer-1 relaying \cite{Layer1Relay}. 
Another key-factor in modern communication systems is the use of multiple antennas. It is established that the use of multiple antennas grants higher data rates and lower bit error rates \cite{TseMIMO}. As a result, recently a great deal of research has focused on MIMO relaying, where a multiple antenna, non-regenerative relay precodes the signal received from the source by an AF matrix, and then forwards it to the destination.

Most previous papers in this research direction consider source and relay precoding matrix allocation for the optimization of traditional performance measures such as achievable rate and minimum mean square error \cite{munoz2007linear,chae2008mimo,Vaze2011,Rong2009,Calvo2009,yu-power,MIMOAFRelayPowerA,WangTao2012,Heath2013,ZapTWC12} and references therein. 
However, the consideration that in mobile networks the nodes are typically battery-powered, thus having a limited lifetime, as well as the concerns for sustainable growth due to the increasing demand for energy, have garnered a great deal of interest on an efficient use of energy in wireless networks \cite{commmag_si,jsac_si} both in academia and in industry. Information and communication technologies (ICT) consume about 2$\%$ of the entire world energy consumption, and the situation is likely to reach the point wherein ICT equipments in large cities will require more energy than it is actually available \cite{Pickavet}. One approach in this sense is to consider the minimization of the transmit power subject to QoS constraint \cite{MultihopMIMORelayRong11, PowerallMIMOAF2012}. However, the trade-off between achieving high data rates and limiting energy consumptions is mathematically more thoroughly described by considering the optimization of new, fractional performance measures, which are measured in bit/Joule and thus naturally represent the efficiency with which each Joule of energy drained from the battery is being used to transmit information. Resource allocation for bit/Joule EE optimization has been extensively analyzed in single-antenna, one-hop networks and several performance metrics have been proposed. In \cite{ZapOFDMA,WidelyLinear,Betz2008,Meshkati2007} and references therein, the EE is defined as the ratio between the achieved throughput and the consumed power. Instead, in \cite{Miao2011,Isheden2011} the ratio between the achievable rate and the consumed power has been considered. As for one-hop, multiple-antenna networks, fewer results are available. In \cite{BuzziMIMO}, the ratio between the throughput and the consumed power is optimized, but the simplifying assumption of single-stream transmission is made. In \cite{belmega2011energy} the EE is defined as the ratio between the goodput and the transmit power and the problem of transmit covariance matrix allocation is studied. 
In \cite{MIMOBroadcastEE} a broadcast MIMO channel is considered and uplink-downlink duality is exploited to come up with a transmit covariance matrix allocation algorithm so as to maximize the ratio between the system capacity and the consumed power. Few results are available for relay-assisted single-antenna networks, too. In \cite{Zap2011Letter} competitive power control algorithms for EE maximization in relay-assisted single-antenna multiple access networks are devised, while \cite{ZapponeTWC} extends the results of \cite{Zap2011Letter} to interference networks. 

All of the previously cited works assume that perfect CSI is available for resource allocation purposes, which might not be feasible in real-world systems. Indeed, a significant research trend is to devise resource allocation algorithms that require only statistical CSI, thus reducing the amount of communication overhead. As far as MIMO systems are concerned, contributions in this direction have mainly focused on the optimization of traditional performance measures such as achievable rate \cite{soysal2007optimum,soysal2009optimality,ZapRelayMIMO2013} and mean square error \cite{TaoImperfect2012}. Instead, less attention has been given to the problem of EE optimization with statistical CSI. In \cite{Chong2011c,AlfanoChong}, a one-hop MIMO link is considered and the transmit covariance matrix is allocated so as to maximize the ratio between the ergodic capacity and the consumed power. Instead, no results are available for bit/Joule EE optimization in relay-assisted MIMO systems, and even the simpler case in which perfect CSI is assumed is an almost unexplored field. Indeed, to the best of our knowledge, the first contribution in this direction is the conference paper \cite{CaoCAMAD12}, where preliminary results on EE maximization in MIMO relay-assisted systems with perfect CSI are provided.

Motivated by this background, this work is aimed at providing a thorough investigation of energy-efficient resource allocation in MIMO relay-assisted systems, with perfect and statistical CSI. The EE is defined as the ratio between the system's achievable rate and the consumed power. In the definition of the consumed power, not only the transmit power, but also the circuit power dissipated in the devices' electronic circuitry is accounted for. In such a scenario, energy-efficient resource allocation algorithms that jointly allocate the source and relay precoding matrix subject to power and QoS constraints have been devised. In particular, the following cases have been considered.
\begin{enumerate}
\item Perfect CSI is available for both source-to-relay and relay-to-destination channel.
\item Perfect CSI is available only for the relay-to-destination channel, while the source-to-relay channel is only statistically known.
\item Perfect CSI is available only for the source-to-relay channel, while the relay-to-destination channel is only statistically known.
\end{enumerate}
In all three cases, the fractional, non-convex optimization problem to be solved has been tackled by means of a two-step approach. First, the optimal source and relay transmit directions have been determined in closed form. Next, plugging the optimal transmit directions in the objective function, it has been shown that the resulting problem is separately pseudo-concave in the source and relay power allocation vectors. Thus, the alternating maximization algorithm coupled with fractional programming tools have been used to complete the resource allocation process. 
Moreover, with reference to scenarios 2) and 3) sufficient conditions for the optimality of source beamforming transmission have been derived, which allows to reduce the complexity of the resource allocation phase. Otherwise stated, sufficient conditions under which the optimal power allocation at the source is to concentrate all the available power on just one data stream have been derived.

The rest of the paper is organized as follows. Section \ref{SystemModel} describes the considered scenario, formally stating the problem to be tackled. In Section \ref{Sec:PerfectCSI} the energy-efficient resource allocation problem is solved assuming perfect CSI is available for both the source-to-relay and the relay-to-destination channel. Section \ref{Sec:Uplink} addresses the case in which only statistical CSI for the source-to-relay channel is available, while Section \ref{Sec:Downlink} tackles the opposite case in which the relay-to-destination channel is statistically known. In Section \ref{Sec:OptBeam} the optimality of beamforming transmission is investigated for the scenarios considered in Sections \ref{Sec:Uplink} and \ref{Sec:Downlink}. 
Numerical results are provided in Section \ref{Sec:NumericalResults}, while concluding remarks are provided in Section \ref{Sec:Conclusion}. Some lemmas which are instrumental to the derivation of the theoretical results are provided in the Appendix.

\textbf{Notation:} In the sequel, $\mathbb{E}[\cdot]$ is the statistical expectation operator, $\bI_n$ denotes an $n\times n$ identity matrix, $(\cdot)^H$, ${\rm tr}(\cdot)$, $|\cdot|$, and $(\cdot)^{+}$ denote Hermitian, trace, determinant, and pseudo-inversion of a matrix, respectively. ${\rm diag}(v_{1},\ldots,v_{N})$ denotes a diagonal matrix with $\{v_{n}\}_{n=1}^{N}$ as diagonal elements. while ${\cal H}^{n}$ denotes the space of $n \times n$, Hermitian, positive semidefinite matrices. Matrix inequalities will be intended in the L{\"o}wner sense\footnote{For any two Hermitian, positive semidefinite matrices $\bM_{1}$ and $\bM_{2}$, $\bM_{1}\succeq\bM_{2}$ means by definition that $\bM_{1}-\bM_{2}$ is positive semidefinite.}. The acronym EVD and SVD stand for eigenvalue decomposition and singular value decomposition, respectively, and, without loss of generality, in all EVDs and SVDs, the eigenvalues and singular values will be assumed to be arranged in decreasing order. 

\section{System Model and Problem Statement}\label{SystemModel}
%Consider a relay-assisted MIMO system. Denote by $N_{S}$, $N_{R}$, and $N_{D}$ the number of antennas equipped at source, at the relay, and at the destination, respectively.
Consider a relay-assisted MIMO system consisting of one source $\mathcal{S}$, one half-duplex AF relay $\mathcal{R}$ and one destination $\mathcal{D}$, which are equipped with $N_S$, $N_R$ and $N_D$ antennas, respectively.
Let $\bs$ be the source's unit-norm symbol vector, and $\bx=\bQ^{1/2}\bs$, with $\bQ=E[\bx\bx^{H}]$ being the source transmit covariance matrix. Let us also denote by $\bH$ and $\bG$ the source-relay and relay-destination channels, and by $\bA$ the AF relay matrix. Then, the signals $\by_{R}$ and $\by_{D}$ received at the relay and destination respectively, can be written as $\by_{R}=\bH\bQ^{1/2}\bs+\bn_{R}$ and $\by_{D}=\bG\bA\bH\bQ^{1/2}\bs+\bG\bA\bn_{R}+\bn_{D}$, with $\bn_{R}$ and $\bn_{D}$ being the thermal noise at relay and destination, modeled as zero-mean complex circular Gaussian vectors  with covariance matrices $\sigma_{R}^{2}\bI_{N_{R}}$ and $\sigma_{D}^{2}\bI_{N_{D}}$, respectively.

The goal of the resource allocation process is to maximize the efficiency with which the system nodes employ the energy supply at their disposal to transmit information. The efficiency of any physical system is usually defined by the benefit-cost ratio, and in communication systems two natural measures of benefit and cost are the achievable rate and the consumed energy. The ratio between the achievable rate in a communication system and the consumed energy is commonly referred to as the global energy efficiency (GEE) of the system. Clearly, a trade-off exists between ensuring high achievable rates and saving as much energy as possible. Therefore, the maximization of the GEE is not trivial and fundamentally different from achievable rate maximization, since the resource allocation algorithm should aim at striking the optimal balance between high data-rates and low consumed energy.

In the considered system, the achievable rate is expressed in bits/s/Hz as \cite{yu-power}
\beq
\small
R(\bQ,\bA)=\frac{1}{2}{\rm log}\left|\bI_{N_{D}}+\bW^{-1}\bG\bA\bH\bQ\bH^{H}\bA^{H}\bG^{H}\right|\;,\label{eq:Rate1}
\eeq
with $\bW=\sigma_{D}^{2}\bI_{N_{D}}+\sigma_{R}^{2}\bG\bA\bA^{H}\bG^{H}$ being the overall noise covariance matrix, and the factor $\frac{1}{2}$ stemming from the fact that the signal vector is transmitted in two time slots. Then, denoting by $T$ the total transmission time, the amount of information that can be reliably transmitted in the time-interval $T$ is $T\cdot R(\mat{Q}, \mat{A})$ bits/Hz, with the source and relay transmit power constraints 
\begin{gather}
\small
P_{S}(\bQ)={\rm tr}(\mat{Q})\leq P_S^{max}\notag \\ 
\label{eq:RP}
\small
P_{R}(\bQ,\bA)={\rm tr}\left(\mat{A}(\mat{H}\mat{Q}\mat{H}^H+ \sigma_R^2 \mat{I}_{N_R})\mat{A}^H\right)\leq P_R^{max}\;,
\end{gather}
wherein $P_S^{max}$ and $P_R^{max}$ denote the maximum feasible transmit powers at $\mathcal{S}$ and $\mathcal{R}$, respectively. 

In a half-duplex relay channel, each node has three operation modes: transmission, reception and idle mode \cite{EEtwowayRelay}. The power consumptions in these modes are denoted by $P/\zeta + P^{ct}$, $P^{cr}$ and $P^{ci}$, respectively, where $P$ is the transmit power, $\zeta \in (0, 1]$ is the power amplifier efficiency, $P^{ct}$, $P^{cr}$, and $P^{ci}$ are the circuit power consumption in transmission, reception, and idle mode, respectively. We assume that $P^{ct}$, $P^{cr}$ and $P^{ci}$ are modeled as constant terms independent of the data rate \cite{EEtwowayRelay}, \cite{RFPc1Cui2005}. 
In the first time slot, $\mathcal{S}$, $\mathcal{R}$ and $\mathcal{D}$ are in transmission mode, reception mode and idle mode, respectively. In the second time slot, $\mathcal{S}$, $\mathcal{R}$ and $\mathcal{D}$ are in idle mode, transmission mode and reception mode, respectively. Then, the amount of energy consumed in the time-interval $T$ can be expressed as
\beq\label{eq:Power1}
\small
E(\mat{Q}, \mat{A}) =\frac{T}{2} \left(\frac{P_S(\mat{Q})}{\zeta_S} +\frac{P_R(\mat{Q}, \mat{A})}{\zeta_R} + P_c \right), 
\eeq
where $P_c= (P_S^{ct} + P_R^{cr} + P_D^{ci} + P_S^{ci} + P_R^{ct} + P_D^{cr})$ is the total circuit power dissipated in the network nodes. For notational ease, and without loss of generality, in the following we assume $\zeta_S=\zeta_R=1$. Then, the GEE is defined as
\beq\label{Eq:GEE}
\small
{\rm GEE}=T\frac{ R(\bQ,\bA)}{E(\bQ,\bA)}\;.
\eeq
Note that (\ref{Eq:GEE}) is measured in $\ds\frac{{\rm bit}/Hz}{J}$, thus representing a natural measure of the efficiency with which each Joule of energy is used. The problem to be tackled is that of GEE maximization subject to the power constraints (\ref{eq:RP}) and to the QoS constraint $R(\bQ,\bA)\geq R_{S}^{min}$, with $R_{S}^{min}$ being the minimum acceptable achievable rate.
Such problem will be addressed with reference to the following scenarios: 
\begin{enumerate}
\item $\mathcal{S}$ and $\mathcal{R}$ have perfect CSI for both channels $\bH$ and $\bG$. 
\item $\mathcal{S}$ and $\mathcal{R}$ have perfect knowledge of the relay-to-destination channel $\bG$, but only statistical CSI for the source-to-relay channel $\bH$.
\item $\mathcal{S}$ and $\mathcal{R}$ have perfect knowledge of the source-to-relay channel $\bH$, but only statistical CSI for the relay-to-destination channel $\bG$.
\end{enumerate}
In all scenarios it is assumed that $\mathcal{D}$ has perfect knowledge of both channels $\bH$ and $\bG$.

\section{GEE maximization with perfect CSI}\label{Sec:PerfectCSI}
Assume both $\mathcal{S}$ and $\mathcal{R}$ have perfect CSI on $\bH$ and $\bG$. For future reference, denote the SVDs of the channels by $\small\mat{H} = \mat{U}_H\mat{\Lambda}_H^{1/2}\mat{V}_H^H$, $\small\mat{G} = \mat{U}_G\mat{\Lambda}_G^{1/2}\mat{V}_{G}^H$, while the EVD of $\bQ$ and SVD of $\bA$ are given by $\bQ=\bU_{Q}\bLambda_{Q}\bU_{Q}^{H}$ and $\bA=\bU_{A}\bLambda_{A}^{1/2}\bV_{A}^{H}$.
The resource allocation problem can be formulated as the maximization problem
\begin{equation}\label{eq:P0}
\small
\begin{array}{lll}
\ds\max_{\mat{Q}\succeq \mat{0},\mat{A}} \frac{{\rm log}\left|\bI_{N_{D}}+\bW^{-1/2}\bG\bA\bH\bQ\bH^{H}\bA^{H}\bG^{H}\bW^{-1/2}\right|}{{\rm tr}(\bQ) +{\rm tr}\left(\mat{A}(\mat{H}\mat{Q}\mat{H}^H+\sigma_R^2 \bI_{N_R})\mat{A}^H\right) + P_c}\;.  \\
\mathrm{s.t.}\;\;\;{\rm log}\left|\bI_{N_{D}}+\bW^{-1/2}\bG\bA\bH\bQ\bH^{H}\bA^{H}\bG^{H}\bW^{-1/2}\right| \geq R_S^{min} \\
{\rm tr}(\bQ) \leq P_{S}^{max}\;,\;\;{\rm tr}\left(\mat{A}(\mat{H}\mat{Q}\mat{H}^H+\sigma_R^2 \bI_{N_R})\mat{A}^H\right) \leq P_{R}^{max}
\end{array}\;.
\end{equation}
Problem (\ref{eq:P0}) is a complex fractional problem which is not jointly convex in $(\mat{Q}, \mat{A})$. It should also be remarked that, while the numerator of the GEE is well-known to be maximized by diagonalizing the channel matrices and arranging the eigenvalues of $\bA\bA^{H}$ and $\bQ$ in decreasing order, the same allocation of $\bA$ and $\bQ$ would actually maximize the denominator, which is instead minimized by arranging the eigenvalues of $\bA\bA^{H}$ and $\bQ$ in increasing order. Therefore, it is not straightforward to conclude that diagonalization is optimal when maximizing the GEE. In order to show that diagonalization is indeed optimal, the following result provides a change of variables that allows to rewrite the GEE as a fraction whose numerator and denominator will be shown to be simultaneously maximized and minimized, respectively, by diagonalization.
\begin{proposition}\label{Prop:OptDirCSI}
Consider Problem (\ref{eq:P0}). The optimal $\bQ$ and $\bA$ are such that $\bU_{Q}=\bV_{H}$, $\bU_{A}=\bV_{G}$, and $\bV_{A}=\bU_{H}$. 
\end{proposition}
\begin{IEEEproof}
We start by rewriting the objective function as
\beq\label{Eq:ObjHQ0}
\small
\begin{split}
&\frac{{\rm log}\left|\sigma_{D}^{2}\bI_{N_{D}}+\bG\bA\left(\sigma_{R}^{2}\bI_{N_{R}}+\bH\bQ\bH^{H}\right)\bA^{H}\bG^{H}\right|}{{\rm tr}\left(\bA(\bH\bQ\bH^{H}+\sigma_{R}^{2}\bI_{N_{R}})\bA^{H}\right) + {\rm tr}(\bQ) + P_c}\\
&-\frac{{\rm log}\left|\sigma_{D}^{2}\bI_{N_{D}}+\sigma_{R}^{2}\bG\bA\bA^{H}\bG^{H}\right|}{{\rm tr}\left(\bA(\bH\bQ\bH^{H}+\sigma_{R}^{2}\bI_{N_{R}})\bA^{H}\right) + {\rm tr}(\bQ) + P_c}
\end{split}
\eeq
Now, defining the variables $\bY=\bH\bQ\bH^{H}$ and $\bX=\bG\bA(\bY+\sigma_{R}^{2}\bI_{N_{R}})^{1/2}$, (\ref{Eq:ObjHQ0}) can be expressed as 
\beq\label{Eq:ObjHQ}
\small
\frac{{\rm log}\left|\sigma_{D}^{2}\bI_{N_{D}}+\bX\bX^{H}\right|-{\rm log}\left|\sigma_{D}^{2}\bI_{N_{D}}+\sigma_{R}^{2}\bX(\bY+\sigma_{R}^{2}\bI_{N_{R}})^{-1}\bX^{H}\right|}{{\rm tr}(\bH^{+}\bY\bH^{H+}) +{\rm tr}\left(\bG^{+}\bX\bX^{H}\bG^{H+}\right) + P_c}
\eeq
Defining by $\bU_{x}\bLambda_{x}^{1/2}\bV_{x}^{H}$ the SVD of $\bX$ and by $\bU_{y}\bLambda_{y}\bU_{y}^{H}$ the EVD of $\bY$, by virtue of Lemma \ref{Lemma:MinTrace} in Appendix, it follows that the first and second summand in the denominator of (\ref{Eq:ObjHQ}) are minimized when $\bU_{y}=\bU_{H}$ and $\bU_{x}=\bU_{G}$, respectively. Moreover, exploiting Lemma \ref{Lemma:MinDet}, it can also be seen that the numerator is maximized for $\bV_{x}=\bU_{y}=\bU_{H}$. Therefore, such choices for $\bU_{x}$, $\bV_{x}$, and $\bU_{y}$ simultaneously maximize the numerator and minimize the denominator of (\ref{Eq:ObjHQ}). Moreover, they are also feasible because the numerator of the objective is also the LHS of the QoS constraint, while the first and second summand in the denominator are the LHS of the power constraints. Next, from the expression of $\bY$ we have $\bY=\bU_{y}\bLambda_{y}\bU_{y}^{H}=\bU_{H}\bLambda_{H}^{1/2}\bV_{H}^{H}\bU_{Q}\bLambda_{Q}\bU_{Q}^{H}\bV_{H}\bLambda_{H}^{1/2}\bU_{H}^{H}$, from which it follows that in order to achieve $\bU_{y}=\bU_{H}$, the relation $\bU_{Q}=\bV_{H}$ needs to hold. Similarly, for $\bX$ we have $\bX=\bU_{x}\bLambda_{x}\bV_{x}^{H}=\bU_{G}\bLambda_{G}^{1/2}\bV_{G}^{H}\bU_{A}\bLambda_{A}^{1/2}\bV_{A}^{H}\bU_{H}(\bLambda_{H}^{1/2}\bLambda_{Q}\bLambda_{H}^{1/2}+\sigma_{R}^{2}\bI_{N_{R}})^{1/2}\bU_{H}^{H}$. Thus, in order to achieve $\bU_{x}=\bU_{G}$ and $\bV_{x}=\bU_{H}$, the relations $\bU_{A}=\bV_{G}$, and $\bV_{A}=\bU_{H}$ need to hold.
\end{IEEEproof}
\begin{remark}\label{Rem:NonFullRank}
In the proof of Proposition \ref{Prop:OptDirCSI} it has been implicitly assumed that both $\bH$ and $\bG$ are tall full-rank matrices. However, this assumption has been made only for notational ease and the result of Proposition \ref{Prop:OptDirCSI} can be readily extended to the case of generic matrices $\bH$ and $\bG$, too. For example, assume $\bG$ and $\bH$ are wide full-rank\footnote{The case of non-full-rank channel matrices is of little practical relevance since $\bH$ and $\bG$ will be full-rank with probability 1. However, the method reported here can be applied also to rank-deficient matrices.} matrices. Thus we have $N_{S}\geq N_{R}\geq N_{D}$ and defining the matrices $\widetilde{\bQ}=\bV_{H}^{H}\bQ\bV_{H}$ and $\widetilde{\bA}=\bV_{G}^{H}\bA$, the objective of (\ref{eq:P0}) can be rewritten as equation (\ref{Eq:ObjRank}),
\setcounter{equation}{8}
\begin{figure*}[!t]
\setcounter{MYtempeqncnt}{\value{equation}}
\setcounter{equation}{7}
\beq\label{Eq:ObjRank}
\small
\frac{{\rm log}\left|\sigma_{D}^{2}\bI_{N_{D}}+\bLambda_{G,N_{D}}^{1/2}\widetilde{\bA}_{N_{D}}(\bU_{H}\bLambda_{H,N_{R}}^{1/2}\widetilde{\bQ}_{N_{R}}\bLambda_{H,N_{R}}^{1/2}\bU_{H}^{H}+\sigma_{R}^{2}\bI_{N_{R}})\widetilde{\bA}_{N_{D}}^{H}\bLambda_{G,N_{D}}^{1/2}\right|-{\rm log}\left|\sigma_{D}^{2}\bI_{N_{D}}+\sigma_{R}^{2}\bLambda_{G,N_{D}}^{1/2}\widetilde{\bA}_{N_{D}}\widetilde{\bA}_{N_{D}}^{H}\bLambda_{G,N_{D}}^{1/2}\right|}{{\rm tr}(\widetilde{\bQ}) +{\rm tr}\left(\widetilde{\mat{A}}(\bU_{H}\bLambda_{H,N_{R}}^{1/2}\widetilde{\bQ}_{N_{R}}\bLambda_{H,N_{R}}^{1/2}\bU_{H}^{H}+\sigma_R^2 \bI_{N_R})\widetilde{\mat{A}}^H\right) + P_c}
\eeq
\hrulefill
\setcounter{equation}{\value{MYtempeqncnt}}
\end{figure*}
wherein $\bLambda_{G,N_{D}}^{1/2}$ is the left $N_{D}\times N_{D}$ diagonal block of $\bLambda_{G}^{1/2}$, $\bLambda_{H,N_{R}}^{1/2}$ is the left $N_{R}\times N_{R}$ diagonal block of $\bLambda_{H}^{1/2}$, $\widetilde{\bQ}_{N_{R}}$ is the upper-left $N_{R}\times N_{R}$ block of $\widetilde{\bQ}$, while $\widetilde{\bA}_{N_{D}}$ is a $N_{D}\times N_{R}$ matrix containing the first $N_{D}$ rows of $\widetilde{\bA}$. Moreover, for any $\widetilde{\bQ}$ and $\widetilde{\bA}$ we have ${\rm tr}(\widetilde{\bQ})\geq {\rm tr}(\widetilde{\bQ}_{N_{R}})$ and ${\rm tr}\left(\widetilde{\mat{A}}(\bU_{H}\bLambda_{H,N_{R}}^{1/2}\widetilde{\bQ}_{N_{R}}\bLambda_{H,N_{R}}^{1/2}\bU_{H}^{H}+\sigma_R^2 \bI_{N_R})\widetilde{\mat{A}}^H\right)\geq {\rm tr}\left(\widetilde{\mat{A}}_{N_{D}}(\bU_{H}\bLambda_{H,N_{R}}^{1/2}\widetilde{\bQ}_{N_{R}}\bLambda_{H,N_{R}}^{1/2}\bU_{H}^{H}+\sigma_R^2 \bI_{N_R})\widetilde{\mat{A}}_{N_{D}}^H\right)$. Therefore, it is seen that the entries of $\widetilde{\bQ}$ and $\widetilde{\bA}$ that are not contained in $\widetilde{\bQ}_{N_{R}}$ and $\widetilde{\bA}_{N_{D}}$ should be set to zero since they do not affect the numerator of (\ref{Eq:ObjRank}) and only increase the consumed power. Therefore, Problem (\ref{eq:P0}) can be recast in terms of only $\widetilde{\bQ}_{N_{R}}$ and $\widetilde{\bA}_{N_{D}}$, and thus can be solved by means of Proposition \ref{Prop:OptDirCSI}. In the sequel of the paper, similarly to Proposition \ref{Prop:OptDirCSI}, some results will implicitly assume tall, full-rank channel matrices. Such assumptions cause no loss of generality since they can be relaxed with similar techniques as shown here for Proposition \ref{Prop:OptDirCSI}.
%A similar approach can be used to tackle the case of rank-deficient $\bH$ and $\bG$. However, such a scenario is of little practical relevance since typically $\bH$ and $\bG$ are full-rank with probability one.
\end{remark}

As a consequence of Proposition \ref{Prop:OptDirCSI}, denoting by $\lambda_{i,G}$, $\lambda_{i,A}$, $\lambda_{i,H}$, and $\lambda_{i,Q}$, the generic $(i,i)$ entry of the matrices $(\bLambda_{G}^{1/2}\bLambda_{G}^{H/2})$, $\bLambda_{A}$, $(\bLambda_{H}^{1/2}\bLambda_{H}^{H/2})$, and $\bLambda_{Q}$, respectively, and by $\blambda_{Q}$ and $\blambda_{A}$ the vectors $\{\lambda_{i,Q}\}_{i=1}^{N_{S}}$ and $\{\lambda_{i,A}\}_{i=1}^{N_{R}}$, Problem (\ref{eq:P0}) can be expressed as 
\begin{equation}\label{eq:P1}
\small
\left\{
\begin{array}{lll}
\ds\max_{\blambda_{Q},\blambda_{A}}\frac{\ds\sum_{i=1}^{N_{S}}{\rm log}\left(1+\frac{\lambda_{i,A}\lambda_{i,Q}\lambda_{i,H}\lambda_{i,G}}{\sigma_{D}^{2}+\sigma_{R}^{2}\lambda_{i,A}\lambda_{i,G}}\right)}{\sum_{i=1}^{N_{S}}\lambda_{i,Q}+\sum_{i=1}^{N_{R}}\lambda_{i,A}(\lambda_{i,H}\lambda_{i,Q}+\sigma_{R}^{2})+P_{c}}
\;.  \\
\mathrm{s.t.}\ds\sum_{i=1}^{N_{S}}{\rm log}\left(1+\frac{\lambda_{i,A}\lambda_{i,Q}\lambda_{i,H}\lambda_{i,G}}{\sigma_{D}^{2}+\sigma_{R}^{2}\lambda_{i,A}\lambda_{i,G}}\right) \geq R_S^{min}\\
\lambda_{i,Q}\geq 0\;\; \forall i=1,\ldots,N_{S}\;,\;\;\lambda_{i,A}\geq 0\;\; \forall i=1,\ldots,N_{R} \\
\sum_{i=1}^{N_{S}}\lambda_{i,Q} \leq P_{S}^{max}\;,\;\;\sum_{i=1}^{N_{R}}\lambda_{i,A}(\lambda_{i,H}\lambda_{i,Q}+\sigma_{R}^{2}) \leq P_{R}^{max}.
\end{array}
\right.
\end{equation}
Problem (\ref{eq:P1}), although being a vector-valued, simpler problem than (\ref{eq:P0}), is still non-convex. However, it can be tackled using the tools of fractional programming and the alternating maximization algorithm \cite{BertsekasNonLinear}, as shown in the following. We start by recalling the following result.
\begin{proposition}
Consider the fractional function $f(x)=\ds\frac{N(x)}{D(x)}$.
If $N(x)$ is a concave function and $D(x)$ is a linear function, then $f(x)$ is a pseudo-concave function. Moreover, consider the  function $F$ defined as
\beq\label{Eq:F(mu)}
\small
F(\mu)=\max_{x}\left\{N(x)-\mu D(x)\right\}\;.
\eeq
$F(\mu)$ is continuous, convex and strictly decreasing, while, for fixed $\mu$, the maxmization problem in (\ref{Eq:F(mu)}) is a strictly convex optimization problem. Moreover, the problem of maximizing $f(x)$ is equivalent to the problem of finding the positive zero of $F(\mu)$ .
\end{proposition}
\begin{IEEEproof}
See \cite{FracProgSS1983,NonlinearFracProg}
\end{IEEEproof}
Thus, a pseudo-concave problem can be solved by finding the zero of the auxiliary function $F(\mu)$. This can be done with a superlinear convergence by means of Dinkelbach's algorithm \cite{NonlinearFracProg}.
%\begin{algorithm}
%\small
%\caption{Dinkelbach Algorithm}
%\label{Alg:Dinkelbach}
%\begin{algorithmic}
%\STATE
%Initialize $\mu_0$ satisfying $F(\mu_0)\geq 0$. Set a tolerance value $\epsilon$ and $n=0$;
%\REPEAT
%\STATE Given $\mu=\mu_n$, find the $x_{n}^{*}$ that solves the convex problem $\ds\max_{x}\left\{N(x)-\mu_{n} D(x)\right\}$;
%\STATE Update $\mu$ as $\mu_{n+1} =\ds \frac{N(x_{n}^{*})}{D(x_{n}^{*})}$;
%\STATE $n=n+1$;
%\UNTIL{$|F(\mu_n)|\leq \epsilon$}
%\end{algorithmic}
%\end{algorithm}
Since pseudo-concave functions have the pleasant property to have no stationary point other than global maximizers \cite{FracProgSS1983}, the output of Dinkelbach's algorithm is guaranteed to be the global solution of the problem, assuming the constraint set of the problem is a convex set. 

Now, it is seen by inspection that the objective of Problem (\ref{eq:P1}) is pseudo-concave in $\blambda_{Q}$ for fixed $\blambda_{A}$ and pseudo-concave in $\blambda_{A}$ for fixed $\blambda_{Q}$. Therefore, one convenient way to solve (\ref{eq:P1}) is to employ the alternating maximization algorithm \cite{BertsekasNonLinear}, according to which Problem (\ref{eq:P1}) can be alternatively solved with respect to $\blambda_{Q}$, for fixed $\blambda_{A}$ and with respect to $\blambda_{A}$, for fixed $\blambda_{Q}$, until the objective converges. Denoting by ${\rm GEE}^{(n)}$ the value of the GEE achieved after the $n$-th iteration of the algorithm, the formal procedure can be stated as follows. 
\begin{algorithm}
\small
\caption{Alternating maximization for Problem (\ref{eq:P1})}
\label{Alg:AOAPerCSI}
\begin{algorithmic}
\STATE
Initialize $\blambda_{Q}^{(0)}$ to a feasible value. Set a tolerance $\epsilon$. Set $n=0$;
\REPEAT
\STATE Given $\blambda_{Q}^{(n)}$, solve Problem (\ref{eq:P1}) with respect to $\blambda_{A}$ to obtain the optimal $\blambda_{A}^{(n+1)}$;
\STATE Given $\blambda_{A}^{(n+1)}$, solve Problem (\ref{eq:P1}) with respect to $\blambda_{Q}$ to obtain the optimal $\blambda_{Q}^{(n+1)}$;
\STATE $n=n+1$;
\UNTIL{$\left|{\rm GEE}^{(n)}-{\rm GEE}^{(n-1)}\right| \leq \epsilon$}
\end{algorithmic}
\end{algorithm}

Convergence of Algorithm \ref{Alg:AOAPerCSI} is ensured by the observation that after each iteration the objective is not decreased and that the objective is upper-bounded. It should also be mentioned that, while the global solution of each subproblem in Algorithm \ref{Alg:AOAPerCSI} is found thanks to Dinkelbach's algorithm, in general it can not be guaranteed that the overall Algorithm \ref{Alg:AOAPerCSI} converges to the global optimum of the GEE because the GEE is not jointly pseudo-concave in $(\blambda_{A},\blambda_{Q})$, and because these two vectors are optimized alternatively. However, if it holds that $\frac{\lambda_{i,A}\lambda_{i,Q}\lambda_{i,H}\lambda_{i,G}}{\sigma_{D}^{2}+\sigma_{R}^{2}\lambda_{i,A}\lambda_{i,G}}>>1$ for all $i=1,\ldots,N_{S}$, then each summand in the numerator of the objective can be approximated by ${\rm log}\left(\frac{\lambda_{i,A}\lambda_{i,Q}\lambda_{i,H}\lambda_{i,G}}{\sigma_{D}^{2}+\sigma_{R}^{2}\lambda_{i,A}\lambda_{i,G}}\right)={\rm log}(\lambda_{i,Q})+{\rm log}\left(\frac{\lambda_{i,A}\lambda_{i,H}\lambda_{i,G}}{\sigma_{D}^{2}+\sigma_{R}^{2}\lambda_{i,A}\lambda_{i,G}}\right)$, which is a strictly jointly concave function of $\blambda_{Q}$ and $\blambda_{A}$. As a consequence, since strictly pseudo-concave functions enjoy the property to have only one stationary point, which is the function's global maximizer, it is likely that Algorithm \ref{Alg:AOAPerCSI} converges to the GEE global maximizer. Indeed, the numerical results that will be presented in Section \ref{Sec:NumericalResults} confirm such conjecture. 
Algorithm \ref{Alg:AOAPerCSI} can be implemented either centrally or in a distributed fashion. In the former case, it could be implemented at the relay, which then feeds back the resulting $\bQ$ to the source. In the latter scenario the algorithm should be run in parallel at ${\cal S}$ and ${\cal R}$, which, at the end, will automatically learn their respective precoding matrices. 

\section{GEE maximization with partial CSI on $\bH$}\label{Sec:Uplink}
Assume that the relay-to-destination channel $\bG$ is perfectly known but that only statistical CSI is available for the source-to-relay channel in the form of covariance feedback. This scenario is realistic in all situations in which the relay-to-destination channel is slowly time-varying, whereas the source-to-relay channel is rapidly time-varying. Indeed, a rapidly varying channel is more difficult to estimate and a resource allocation that depends on such an estimate would have to be updated very frequently, which results in a significant amount of overhead. A typical example is the uplink of a communication system, in which the relay and destination are usually fixed, while the source is a mobile terminal.

Specifically, in this section the channel matrix $\bH$ is expressed according to the Kronecker model \cite{shiu2000fading}, as
\begin{equation}\label{CovarianceH}
\small
\bH=\bR_{r,H}^{1/2}\bZ_{H}\bR_{t,H}^{1/2}\;,
\end{equation}
where $\bZ_{H}$ is a random matrix with independent, zero-mean, unit-variance, proper complex Gaussian entries, whereas $\bR_{r,H}$ and $\bR_{t,H}$ are the positive semidefinite receive and transmit correlation matrices associated to $\bH$. The matrices $\bR_{r,H}$ and $\bR_{t,H}$ are assumed known whereas the matrix $\bZ_{H}$ is unknown at the source and relay.
The covariance feedback model has been widely used in the literature, \cite{soysal2009optimality,soysal2007optimum}, \cite{jafar2004transmitter,jorswieck2004channel}, and applies for example to scenarios in which relay and base station are surrounded by local scatterers that induce the matrices $\bR_{t,H}$ and $\bR_{r,H}$, and are separated by a rich multipath environment that is modeled by the matrix $\bZ_{H}$.
We also remark that by letting the transmit and receive correlation matrices be identity matrices, the special notable case in which $\bH$ is completely unknown and modeled as a random matrix with independent, zero-mean, unit-variance, proper complex Gaussian entries is obtained. For future reference, let us define
\beq\label{Eq:EVD_H}
\small
\bR_{r,H}^{1/2}=\bU_{r,H}\bLambda_{r,H}^{1/2}\bU_{r,H}^{H}\;,\;\;
\bR_{t,H}^{1/2}=\bU_{t,H}\bLambda_{t,H}^{1/2}\bU_{t,H}^{H}\;.
\eeq
As for the performance measure to optimize, since only statistical knowledge of $\bH$ is available, it is not possible to optimize the instantaneous GEE (\ref{Eq:GEE}). Instead, the GEE of the considered system should be defined recalling the original definition of the GEE which is the ratio between the benefit and cost of the system. For the case at hand, the benefit is given by the ergodic achievable rate, while the cost is the average consumed energy, which leads to the definition
\beq\label{Eq:AvGEE}
\small
{\rm GEE}=T\frac{\mathbb{E}_{Z_{H}}\left[R(\bQ,\bA)\right]}{\mathbb{E}_{Z_{H}}\left[E(\bQ,\bA)\right]}\;.
\eeq
It should be mentioned that another approach would be to consider the maximization of the average of (\ref{Eq:GEE}) with respect to $\bZ_{H}$, namely
\beq\label{Eq:AvGEE1}
\small
{\widetilde{\rm GEE}}=T\mathbb{E}_{Z_{H}}\left[\frac{R(\bQ,\bA)}{E(\bQ,\bA)}\right]\;.
\eeq
However, (\ref{Eq:AvGEE1}) can not be considered a proper GEE since it is not the ratio between the benefit produced by the system and the cost incurred to achieve such benefit. Thus, (\ref{Eq:AvGEE1}) does not represent the efficiency with which the resources are being used to produce the necessary goods, as instead does (\ref{Eq:AvGEE}). Therefore, (\ref{Eq:AvGEE1}) will not be considered as performance measure and the focus will be on (\ref{Eq:AvGEE}). The optimization problem at hand can be formulated as follows
\begin{equation}\label{Prob:JointProbH}
\small
\begin{array}{lll}
\ds\max_{\bQ,\bA}\ds\frac{\mathbb{E}_{Z_{H}}\left[{\rm log}\left(\frac{\left|\sigma_{D}^{2}\bI_{N_{D}}+\bG\bA\left(\bH\bQ\bH^{H}+\sigma_{R}^{2}\bI_{N_{R}}\right)\bA^{H}\bG^{H}\right|}{\left|\sigma_{D}^{2}\bI_{N_{D}}+\sigma_{R}^{2}\bG\bA\bA^{H}\bG^{H}\right|}\right)\right]}{\mathbb{E}_{Z_{H}}\left[{\rm tr}(\bA(\bH\bQ\bH^{H}+\sigma_{R}^{2}\bI_{N_{R}})\bA^{H})\right]+{\rm tr}(\bQ)+P_{c}}
\\
{\rm s.t.\quad}\mathbb{E}_{Z_{H}}\left[{\rm tr}(\bA(\bH\bQ\bH^{H}+\sigma_{R}^{2}\bI_{N_{R}})\bA^{H})\right]\leq P_{R}^{max}\\
{\rm tr}(\bQ)\leq P_{S}^{max} \;,\;\bQ\succeq 0\\
\mathbb{E}_{Z_{H}}\left[{\rm log}\left(\frac{\left|\sigma_{D}^{2}\bI_{N_{D}}+\bG\bA\left(\bH\bQ\bH^{H}+\sigma_{R}^{2}\bI_{N_{R}}\right)\bA^{H}\bG^{H}\right|}{\left|\sigma_{D}^{2}\bI_{N_{D}}+\sigma_{R}^{2}\bG\bA\bA^{H}\bG^{H}\right|}\right)\right]\geq R_{S}^{min}
\end{array}
\;.
\end{equation}
The following proposition determines the optimal source eigenvector matrix $\bU_{Q}$.
\begin{proposition}\label{Prop:SourceH}
Consider Problem (\ref{Prob:JointProbH}). For any AF matrix $\bA$, the optimal $\bQ$ is such that $\bU_{Q}=\bU_{t,H}$.
\end{proposition}
\begin{IEEEproof}
To begin with, let us rewrite the objective as
\beq\label{Eq:ObjQ1H}
\small
\frac{\mathbb{E}_{Z_{H}}\left[{\rm log}\left|\bI_{N_{D}}+\bW^{-1/2}\bG\bA\bH\bQ\bH^{H}\bA^{H}\bG^{H}\bW^{-1/2}\right|\right]}{\mathbb{E}_{Z_{H}}\left[{\rm tr}(\bA\bH\bQ\bH^{H}\bA^{H})\right]+{\rm tr}(\bQ)+\sigma_{R}^{2}{\rm tr}(\bA\bA^{H})+P_{c}}\;.
\eeq
Next, plugging (\ref{CovarianceH}) and (\ref{Eq:EVD_H}), (\ref{Eq:ObjQ1H}) can be expressed as in (\ref{Eq:ObjQ2H}), 
\setcounter{equation}{17}
\begin{figure*}[!t]
\setcounter{MYtempeqncnt}{\value{equation}}
\setcounter{equation}{16}
\beq\label{Eq:ObjQ2H}
\small
\frac{\mathbb{E}_{Z_{H}}\left[{\rm log}\left|\bI_{N_{D}}+\bW^{-1/2}\bG\bA\bR_{r,H}^{1/2}\bZ_{H}\bLambda_{t,H}^{1/2}\bU_{t,H}^{H}\bU_{Q}\bLambda_{Q}\bU_{Q}^{H}\bU_{t,H}\bLambda_{t,H}^{1/2}\bZ_{H}^{H}\bR_{r,H}^{1/2}\bA^{H}\bG^{H}\bW^{-1/2}\right|\right]}{\mathbb{E}_{Z_{H}}\left[{\rm tr}(\bA\bR_{r,H}^{1/2}\bZ_{H}\bLambda_{t,H}^{1/2}\bU_{t,H}^{H}\bU_{Q}\bLambda_{Q}\bU_{Q}^{H}\bU_{t,H}\bLambda_{t,H}^{1/2}\bZ_{H}^{H}\bR_{r,H}^{1/2}\bA^{H})\right]+{\rm tr}(\bLambda_{Q})+\sigma_{R}^{2}{\rm tr}(\bLambda_{A})+P_{c}}
\eeq
\hrulefill
\setcounter{equation}{\value{MYtempeqncnt}}
\end{figure*}
where it has been exploited the fact that multiplying $\bZ_{H}$, from left or right, by a unitary matrix does not change its distribution.
Next, defining $\bX=\bU_{t,H}^{H}\bU_{Q}\bLambda_{Q}\bU_{Q}^{H}\bU_{t,H}$, the numerator of (\ref{Eq:ObjQ2H}) can be written as the concave function $N_{Q}(\bX)$, shown in (\ref{Eq:NumObjQ2H}).
\setcounter{equation}{18}
\begin{figure*}[!t]
\setcounter{MYtempeqncnt}{\value{equation}}
\setcounter{equation}{17}
\beq\label{Eq:NumObjQ2H}
\small
N_{Q}(\bX)=\mathbb{E}_{Z_{H}}\left[{\rm log}\left|\bI_{N_{D}}+\bW^{-1/2}\bG\bA\bR_{r,H}^{1/2}\bZ_{H}\bLambda_{t,H}^{1/2}\bX\bLambda_{t,H}^{1/2}\bZ_{H}^{H}\bR_{r,H}^{1/2}\bA^{H}\bG^{H}\bW^{-1/2}\right|\right]
\eeq
\hrulefill
\setcounter{equation}{\value{MYtempeqncnt}}
\end{figure*}
At this point, defining $\bGamma$ as in the proof of Lemma \ref{Lemma:DiagonalX} in Appendix, we have (\ref{Eq:PropHQStep1}), where it has been exploited that $\bGamma$ and $\bLambda_{t,H}^{1/2}$ commute because they are both square diagonal matrices, and that $\bGamma$ is a unitary matrix and thus the random matrix $\widetilde{\bZ}_{H}=\bZ_{H}\bGamma$ has the same distribution as $\bZ_{H}$.
\setcounter{equation}{19}
\begin{figure*}[!t]
\setcounter{MYtempeqncnt}{\value{equation}}
\setcounter{equation}{18}
\beq\label{Eq:PropHQStep1}
\small
\begin{split}
N_{Q}(\bGamma\bX\bGamma)&=\mathbb{E}_{Z_{H}}\left[{\rm log}\left|\bI_{N_{D}}+\bW^{-1/2}\bG\bA\bR_{r,H}^{1/2}\bZ_{H}\bLambda_{t,H}^{1/2}\bGamma\bX\bGamma\bLambda_{t,H}^{1/2}\bZ_{H}^{H}\bR_{r,H}^{1/2}\bA^{H}\bG^{H}\bW^{-1/2}\right|\right]\\
&=\mathbb{E}_{Z_{H}}\left[{\rm log}\left|\bI_{N_{D}}+\bW^{-1/2}\bG\bA\bR_{r,H}^{1/2}\bZ_{H}\bGamma\bLambda_{t,H}^{1/2}\bX\bLambda_{t,H}^{1/2}\bGamma\bZ_{H}^{H}\bR_{r,H}^{1/2}\bA^{H}\bG^{H}\bW^{-1/2}\right|\right]\\
&=\mathbb{E}_{Z_{H}}\left[{\rm log}\left|\bI_{N_{D}}+\bW^{-1/2}\bG\bA\bR_{r,H}^{1/2}\bZ_{H}\bLambda_{t,H}^{1/2}\bX\bLambda_{t,H}^{1/2}\bZ_{H}^{H}\bR_{r,H}^{1/2}\bA^{H}\bG^{H}\bW^{-1/2}\right|\right]=N_{Q}(\bX)
\end{split}
\eeq
\hrulefill
\setcounter{equation}{\value{MYtempeqncnt}}
\end{figure*}
Hence, by virtue of Lemma \ref{Lemma:DiagonalX}, it holds that $N_{Q}(\bX)$ is maximized when $\bX$ is diagonal.  Next, we show that this choice for $\bU_{Q}$ is also optimal as far as the denominator of (\ref{Eq:ObjQ2H}) is concerned. Indeed, the part of the denominator of (\ref{Eq:ObjQ2H}) that depends on $\bU_{Q}$ is the function
\beq
\small
D_{Q}(\bX)=\mathbb{E}_{Z_{H}}\left[{\rm tr}(\bA\bR_{r,H}^{1/2}\bZ_{H}\bLambda_{t,H}^{1/2}\bX\bLambda_{t,H}^{1/2}\bZ_{H}^{H}\bR_{r,H}^{1/2}\bA^{H})\right]\;,
\eeq
which is linear in $\bX=\bU_{x}\bLambda_{x}\bU_{x}^{H}$. Moreover, it is easy to check that $D_{Q}(\bGamma\bX\bGamma)=D_{Q}(\bX)$. Thus, employing again Lemma \ref{Lemma:DiagonalX}, it follows that we can set $\bU_{x}=\bI_{N_{T}}$ without affecting 
$D_{Q}(\bX)$. Moreover, this choice is also feasible since it maximizes the LHS of the QoS constraint, while leaving unaffected the LHS of the power constraints. Finally, from $\bU_{x}=\bI_{N_{T}}$ we obtain $\bU_{Q}=\bU_{t,H}$.
\end{IEEEproof}
Next, we tackle the optimization with respect to the left and right eigenvector matrices of $\bA$.
\begin{proposition}\label{Prop:RelayH}
Consider Problem (\ref{Prob:JointProbH}). For any source covariance matrix $\bQ$ with the optimal structure $\bU_{Q}=\bU_{t,H}$, the optimal $\bA$ is such that $\bU_{A}=\bV_{G}$ and $\bV_{A}=\bU_{r,H}$, if either $\bLambda_{G}$ or $\bLambda_{r,H}$ is a scaled identity matrix\footnote{The proof also holds under more general assumptions as explained next.}. 
\end{proposition}
\begin{IEEEproof}
Plugging the optimal $\bU_{Q}$ into (\ref{Eq:ObjQ2H}) and defining by $\bz_{i,H}$ the $i$-th column of $\bZ_{H}$ for all $i=1,\ldots,N_{S}$, the statistical mean at the denominator can be computed in closed-form as follows.
\beq
\small
\begin{split}
&\mathbb{E}_{Z_{H}}\left[{\rm tr}(\bZ_{H}\bLambda_{t,H}\bLambda_{Q}\bZ_{H}^{H}\bLambda_{r,H}^{1/2}\bU_{r,H}^{H}\bA^{H}\bA\bU_{r,H}\bLambda_{r,H}^{1/2})\right]\\
&={\rm tr}\left(\mathbb{E}_{Z_{H}}\left[\bZ_{H}\bLambda_{t,H}\bLambda_{Q}\bZ_{H}^{H}\right]\bLambda_{r,H}^{1/2}\bU_{r,H}^{H}\bA^{H}\bA\bU_{r,H}\bLambda_{r,H}^{1/2}\right)\\
&={\rm tr}\left(\sum_{i=1}^{N_{S}}\lambda_{i,Q}\lambda_{i,H}^{t}\mathbb{E}_{z_{i,H}}\left[\bz_{i,H}\bz_{i,H}^{H}\right]\bLambda_{r,H}^{1/2}\bU_{r,H}^{H}\bA^{H}\bA\bU_{r,H}\bLambda_{r,H}^{1/2}\right)\\
\label{Eq:DenMean}
&={\rm tr}\left(\bLambda_{Q}\bLambda_{t,H}\right){\rm tr}\left(\bLambda_{r,H}^{1/2}\bU_{r,H}^{H}\bA^{H}\bA\bU_{r,H}\bLambda_{r,H}^{1/2}\right)\;.
\end{split}
\eeq
Next, exploiting again that multiplying $\bZ_{H}$ from left or right by a unitary matrix does not change its distribution, and defining the auxiliary variable $\bY=\bG\bA\bU_{r,H}\bLambda_{r,H}^{1/2}=\bU_{y}\bLambda_{y}^{1/2}\bV_{y}^{H}$, after some elaborations the objective can be expressed as in (\ref{Eq:ObjQ4H}).
\setcounter{equation}{22}
\begin{figure*}[!t]
\setcounter{MYtempeqncnt}{\value{equation}}
\setcounter{equation}{21}
\beq\label{Eq:ObjQ4H}
\small
\frac{\mathbb{E}_{Z_{H}}\left[{\rm log}\left|\bI_{N_{S}}+\bLambda_{Q}\bLambda_{t,H}\bZ_{H}^{H}\bLambda_{y}^{H/2}\left(\sigma_{D}^{2}\bI_{N_{D}}+\sigma_{R}^{2}\bLambda_{y}^{1/2}\bV_{y}^{H}\bLambda_{r,H}^{-1}\bV_{y}\bLambda_{y}^{H/2}\right)^{-1}\bLambda_{y}^{1/2}\bZ_{H}\right|\right]}{{\rm tr}\left(\bLambda_{Q}\bLambda_{t,H}\right){\rm tr}\left(\bY^{H}\bG^{+H}\bG^{+}\bY\right)+\sigma_{R}^{2}{\rm tr}\left(\bLambda_{r,H}^{-1}\bY^{H}\bG^{+H}\bG^{+}\bY\right)+{\rm tr}(\bLambda_{Q})+P_{c}}
\eeq
\hrulefill
\setcounter{equation}{\value{MYtempeqncnt}}
\end{figure*}
Now, defining the matrix $\bY_{v}=\bV_{y}^{H}\bLambda_{r,H}\bV_{y}$, the numerator of (\ref{Eq:ObjQ4H}) is written as the function $N_{A}(\bY_{v})$ in (\ref{Eq:NumObjQ4H}), 
\setcounter{equation}{23}
\begin{figure*}[!t]
\setcounter{MYtempeqncnt}{\value{equation}}
\setcounter{equation}{22}
\beq\label{Eq:NumObjQ4H}
\small
N_{A}(\bY_{v})=\mathbb{E}_{Z_{H}}\left[{\rm log}\left|\bI_{N_{S}}+\bLambda_{Q}\bLambda_{t,H}\bZ_{H}^{H}\bLambda_{y}^{H/2}\left(\sigma_{D}^{2}\bI_{N_{D}}+\sigma_{R}^{2}\bLambda_{y}^{1/2}\bY_{v}^{-1}\bLambda_{y}^{H/2}\right)^{-1}\bLambda_{y}^{1/2}\bZ_{H}\right|\right]
\eeq
\hrulefill
\setcounter{equation}{\value{MYtempeqncnt}}
\end{figure*}
which is concave in $\bY_{v}$ by virtue of Lemma \ref{Lemma:ConcInv} and because of the concavity and monotonicity of the function $\mathbb{E}_{Z_{H}}\left[{\rm log}\left|\cdot\right|\right]$. Moreover, consider a generic $N_{R}\times N_{R}$ matrix $\bGamma$ as in the proof of Lemma \ref{Lemma:DiagonalX}, and also define the $N_{D}\times N_{D}$ matrix\footnote{Here we are assuming $N_{D}\geq N_{R}$, but the proof can be extended to the case $N_{R}<N_{D}$ with similar arguments as those used in Remark \ref{Rem:NonFullRank}, and in this case $\bGamma_{N_{D}}={\rm diag}(\bGamma(1,1),\ldots,\bGamma(N_{D},N_{D}))$.} $\bGamma_{N_{D}}={\rm diag}(\bGamma(1,1),\ldots,\bGamma(N_{R},N_{R}),0,\ldots,0)$. Then, it holds that $N_{A}(\bGamma\bY_{v}\bGamma)=N_{A}(\bY_{v})$, as shown in (\ref{Eq:RelayHStep1}).
\setcounter{equation}{24}
\begin{figure*}[!t]
\setcounter{MYtempeqncnt}{\value{equation}}
\setcounter{equation}{23}
\beq\label{Eq:RelayHStep1}
\small
\begin{split}
N_{A}(\bGamma\bY_{v}\bGamma)&=\mathbb{E}_{Z_{H}}\left[{\rm log}\left|\bI_{N_{S}}+\bLambda_{Q}\bLambda_{t,H}\bZ_{H}^{H}\bLambda_{y}^{H/2}\left(\sigma_{D}^{2}\bI_{N_{D}}+\sigma_{R}^{2}\bLambda_{y}^{1/2}\bGamma\bY_{v}^{-1}\bGamma\bLambda_{y}^{H/2}\right)^{-1}\bLambda_{y}^{1/2}\bZ_{H}\right|\right]\\
&=\mathbb{E}_{Z_{H}}\left[{\rm log}\left|\bI_{N_{S}}+\bLambda_{Q}\bLambda_{t,H}\bZ_{H}^{H}\bLambda_{y}^{H/2}\left(\sigma_{D}^{2}\bI_{N_{D}}+\sigma_{R}^{2}\bGamma_{N_{D}}\bLambda_{y}^{1/2}\bY_{v}^{-1}\bLambda_{y}^{H/2}\bGamma_{N_{D}}\right)^{-1}\bLambda_{y}^{1/2}\bZ_{H}\right|\right]\\
&=\mathbb{E}_{Z_{H}}\left[{\rm log}\left|\bI_{N_{S}}+\bLambda_{Q}\bLambda_{t,H}\bZ_{H}^{H}\bGamma\bLambda_{y}^{H/2}\left(\sigma_{D}^{2}\bI_{N_{D}}+\sigma_{R}^{2}\bLambda_{y}^{1/2}\bY_{v}^{-1}\bLambda_{y}^{H/2}\right)^{-1}\bLambda_{y}^{1/2}\bGamma\bZ_{H}\right|\right]\\
&=\mathbb{E}_{Z_{H}}\left[{\rm log}\left|\bI_{N_{S}}+\bLambda_{Q}\bLambda_{t,H}\bZ_{H}^{H}\bLambda_{y}^{H/2}\left(\sigma_{D}^{2}\bI_{N_{D}}+\sigma_{R}^{2}\bLambda_{y}^{1/2}\bY_{v}^{-1}\bLambda_{y}^{H/2}\right)^{-1}\bLambda_{y}^{1/2}\bZ_{H}\right|\right]=N_{A}(\bY_{v})
\end{split}
\eeq
\hrulefill
\setcounter{equation}{\value{MYtempeqncnt}}
\end{figure*}
There, the first equality follows by observing that $\bLambda_{y}^{1/2}\bGamma=\bGamma_{N_{D}}\bLambda_{y}^{1/2}$, for all $N_{D}\times N_{R}$ pseudodiagonal matrices $\bLambda_{y}^{1/2}$,
while the second equality holds upon noticing that $\bLambda_{y}^{H/2}\left(\sigma_{D}^{2}\bI_{N_{D}}+\sigma_{R}^{2}\bGamma_{N_{D}}\bLambda_{y}^{1/2}\bY_{v}^{-1}\bLambda_{y}^{H/2}\bGamma_{N_{D}}\right)^{-1}\bLambda_{y}^{1/2}=\bGamma\bLambda_{y}^{H/2}\left(\sigma_{D}^{2}\bI_{N_{D}}+\sigma_{R}^{2}\bLambda_{y}^{1/2}\bY_{v}^{-1}\bLambda_{y}^{H/2}\right)^{-1}\bLambda_{y}^{1/2}\bGamma$ for all $N_{D}\times N_{R}$ pseudodiagonal matrices $\bLambda_{y}^{1/2}$ and $N_{R}\times N_{R}$ Hermitian, positive definite matrices $\bY_{v}$. Finally, the third equality stems from the fact that $\bGamma\bZ_{H}$ has the same distribution as $\bZ_{H}$. Then, by virtue of Lemma \ref{Lemma:DiagonalX}, $N_{A}(\bY_{v})$ is maximized for a diagonal $\bY_{v}$, which implies $\bV_{y}=\bI_{N_{R}}$.
As for the denominator of (\ref{Eq:ObjQ4H}), exploiting Lemma \ref{Lemma:MinTrace} it follows that it is minimized\footnote{In fact, this is true also when neither $\bLambda_{G}$ nor $\bLambda_{r,H}$ are scaled identity matrices, but it simultaneously happens that the singular values of ${\bf Y}$ and of ${\bf G}^{+}{\bf Y}$ are both ordered in decreasing order.} with respect to $\bY$ when $\bU_{y}=\bU_{G}$. Thus, finally we have $\bU_{A}=\bV_{G}$, $\bV_{A}=\bU_{r,H}$.
Moreover, these choices are also feasible because, up to the constant (with respect to the optimization variables) term ${\rm tr}(\bLambda_{Q})+P_{c}$, the denominator of (\ref{Eq:ObjQ4H}) is also the LHS of the relay power constraint and the numerator is the LHS of the QoS constraint. 
\end{IEEEproof}
Summing up, after optimizing with respect to $\bU_{Q}$, $\bU_{A}$, and $\bV_{A}$, defining the matrix $\widetilde{\bLambda}_{y}=\bLambda_{y}^{H/2}\bLambda_{y}^{1/2}$, (\ref{Prob:JointProbH}) can be recast as in (\ref{Prob:JointProbH2}),
\setcounter{equation}{25}
\begin{figure*}[!t]
\setcounter{MYtempeqncnt}{\value{equation}}
\setcounter{equation}{24}
\begin{equation}\label{Prob:JointProbH2}
\small
\left\{
\begin{array}{lll}
\ds\max_{\bLambda_{Q},\widetilde{\bLambda}_{y}}\ds\frac{\mathbb{E}_{Z_{H}}\left[{\rm log}\left|\bI_{N_{S}}+\bLambda_{Q}\bLambda_{t,H}\bZ_{H}^{H}\widetilde{\bLambda}_{y}\left(\sigma_{D}^{2}\bI_{N_{R}}+\sigma_{R}^{2}\widetilde{\bLambda}_{y}\bLambda_{r,H}^{-1}\right)^{-1}\bZ_{H}\right|\right]}{{\rm tr}\left(\bLambda_{Q}\bLambda_{t,H}\right){\rm tr}\left(\widetilde{\bLambda}_{y}\widetilde{\bLambda}_{G}\right)+\sigma_{R}^{2}{\rm tr}\left(\widetilde{\bLambda}_{y}\bLambda_{r,H}^{-1}\widetilde{\bLambda}_{G}\right)+{\rm tr}(\bLambda_{Q})+P_{c}}
\\
{\rm s.t.}\quad{\rm tr}\left(\bLambda_{Q}\bLambda_{t,H}\right){\rm tr}\left(\widetilde{\bLambda}_{y}\widetilde{\bLambda}_{G}\right)+\sigma_{R}^{2}{\rm tr}\left(\widetilde{\bLambda}_{y}\bLambda_{r,H}^{-1}\widetilde{\bLambda}_{G}\right)\leq P_{R}^{max}\\
{\rm tr}(\bLambda_{Q})\leq P_{S}^{max} \;,\;\bLambda_{Q}\succeq 0 \;,\;\widetilde{\bLambda}_{y} \succeq 0\\
\mathbb{E}_{Z_{H}}\left[{\rm log}\left|\bI_{N_{S}}+\bLambda_{Q}\bLambda_{t,H}\bZ_{H}^{H}\widetilde{\bLambda}_{y}\left(\sigma_{D}^{2}\bI_{N_{R}}+\sigma_{R}^{2}\widetilde{\bLambda}_{y}\bLambda_{r,H}^{-1}\right)^{-1}\bZ_{H}\right|\right]\geq R_{S}^{min}
\end{array}
\right. 
\end{equation}
\hrulefill
\setcounter{equation}{\value{MYtempeqncnt}}
\end{figure*}
where the $N_{R}\times N_{R}$ matrix $\widetilde{\bLambda}_{G}$ contains the upper-left $N_{R}\times N_{R}$ block of $(\bLambda_{G}^{1/2})^{+H}(\bLambda_{G}^{1/2})^{+}$.
Unfortunately, the objective of Problem (\ref{Prob:JointProbH2}) is neither jointly concave nor jointly pseudo-concave in ($\bLambda_{Q},\widetilde{\bLambda}_{y}$). However, similarly to the perfect CSI scenario, it can be shown that it is separately pseudo-concave in $\bLambda_{Q}$ for fixed $\widetilde{\bLambda}_{y}$ and pseudo-concave in $\widetilde{\bLambda}_{y}$ for fixed $\bLambda_{Q}$. The pseudo-concavity with respect to $\bLambda_{Q}$ for fixed $\widetilde{\bLambda}_{y}$ is apparent, whereas for the pseudo-concavity with respect to $\widetilde{\bLambda}_{y}$ for fixed $\bLambda_{Q}$ Lemma \ref{Lemma:ConcF} in the Appendix is needed. By virtue of Lemma \ref{Lemma:ConcF} and by the concavity and monotonicity of the $E_{Z_{H}}[{\rm log}|\cdot|]$ function, it follows that the numerator of the objective, which coincides also with LHS of the QoS constraint, is concave in $\widetilde{\bLambda}_{y}$, whereas by inspection it can be seen that the denominator of the objective and the LHS of the relay transmit power constraint are linear in $\widetilde{\bLambda}_{y}$. Thus, the pseudo-concavity with respect to $\widetilde{\bLambda}_{y}$ follows and a convenient way to solve (\ref{Prob:JointProbH2}) is again the alternating maximization algorithm. The formal procedure can be devised as follows.
\begin{algorithm}
\small
\caption{Alternating maximization for Problem (\ref{Prob:JointProbH2})}
\label{Alg:AltMaxSCSI}
\begin{algorithmic}
\STATE
Initialize $\bLambda_{Q}^{(0)}$ to a feasible value. Set a tolerance $\epsilon$. Set $n=0$;
\REPEAT
\STATE Given $\bLambda_{Q}^{(n)}$, solve Problem (\ref{Prob:JointProbH2}) with respect to $\widetilde{\bLambda}_{y}$ to obtain the optimal $\widetilde{\bLambda}_{y}^{(n+1)}$;
\STATE Given $\widetilde{\bLambda}_{y}^{(n+1)}$, solve Problem (\ref{Prob:JointProbH2}) with respect to $\bLambda_{Q}$ to obtain the optimal $\bLambda_{Q}^{(n+1)}$;
\STATE $n=n+1$;
\UNTIL{$\left|{\rm GEE}^{(n)}-{\rm GEE}^{(n)}\right| \leq \epsilon$}
\end{algorithmic}
\end{algorithm}

Algorithm \ref{Alg:AltMaxSCSI} enjoys similar properties as Algorithm \ref{Alg:AOAPerCSI}. It is guaranteed to converge since the GEE is upper-bounded, it only requires the solution of pseudo-concave problems, and can be implemented either centrally or in a distributed way. A difference with respect to Algorithm \ref{Alg:AOAPerCSI} is that Algorithm \ref{Alg:AltMaxSCSI} requires the evaluation of statistical expectations, due to the lack of perfect knowledge of the channel. In most applications, this can be easily implemented by numerically evaluating the statistical expectations. Moreover, for those applications in which computational complexity is a critical issue, we remark that the proposed method can still be used by approximating the expectation with a deterministic function. To this end, several approaches exist in the literature to come up with suitable approximations of the $\mathbb{E}_{Z_{H}}\left[{\rm log}|\cdot|\right]$ function that would allow the computation of the expectations in (\ref{Prob:JointProbH2}) in closed-form. For example, the well-known WMMSE algorithm \cite{LuoWMMSE,Heath2013} can be used to replace the rate function at the numerator of the objective by a sum of weighted traces, thus allowing the computation of the expectations by exchanging the statistical mean and trace operators. A second approach is to employ Jensen's inequality to come up with an approximation of the objective function. Discussing such techniques in detail is not the main purpose of this work, but in the following we briefly hint at how Jensen's inequality can be used to devise a sub-optimal resource allocation algorithm that does not require the evaluation of any statistical expectation. By virtue of Jensen's inequality the numerator of the objective of (\ref{Prob:JointProbH2}) can be lower-bounded by 
\beq\label{Eq:JensenNum}
\small{\rm log}\left|\bI_{N_{R}}+{\rm tr}(\bLambda_{Q}\bLambda_{t,H})\widetilde{\bLambda}_{y}\left(\sigma_{D}^{2}\bI_{N_{D}}+\sigma_{R}^{2}\widetilde{\bLambda}_{y}\bLambda_{r,H}^{-1}\right)^{-1}\right|\;,
\eeq
which is still concave in $\widetilde{\bLambda}_{y}$ for fixed $\bLambda_{Q}$. Moreover, it is also concave in $\bLambda_{Q}$, for fixed $\widetilde{\bLambda}_{y}$, since ${\rm tr}(\bLambda_{Q}\bLambda_{t,H})$ is linear in $\bLambda_{Q}$ and the composition of a concave function with a linear one is known to be concave. Then, an alternating maximization algorithm that does not require the evaluation of any statistical expectation can be devised following the same pseudo-code as in Algorithm \ref{Alg:AltMaxSCSI}, but replacing the numerator of the objective of Problem (\ref{Prob:JointProbH2}) with its deterministic approximation (\ref{Eq:JensenNum}).

\section{GEE maximization with partial CSI on $\bG$}\label{Sec:Downlink}
In this section, the opposite case of Section \ref{Sec:Uplink} is considered. The relay-to-destination channel will be assumed only statistically known and expressed according to the Kronecker model, whereas the source-to-relay channel will be assumed perfectly known. For the same reasons explained in Section \ref{Sec:Uplink}, considering this scenario is helpful in all situations in which the source-to-relay channel is slowly time-varying, whereas the relay-to-destination is rapidly time-varying. A typical example is the downlink of a communication system, because in this case source and relay are typically fixed, while the destination is a mobile terminal.

Specifically, in this section the channel $\bG$ is expressed as 
\begin{equation}\label{CovarianceG}
\small
\bG=\bR_{r,G}^{1/2}\bZ_{G}\bR_{t,G}^{1/2}
\end{equation}
where $\bZ_{G}$ is a random matrix with independent, zero-mean, unit-variance, proper complex Gaussian entries, whereas $\bR_{r,G}$ and $\bR_{t,G}$ are the positive semidefinite receive and transmit correlation matrices associated to $\bG$. The matrices $\bR_{r,G}^{1/2}=\bU_{r,G}\bLambda_{r,G}^{1/2}\bU_{r,G}^{H}$ and $\bR_{t,G}^{1/2}=\bU_{t,G}\bLambda_{t,G}^{1/2}\bU_{t,G}^{H}$ are assumed known whereas the matrix $\bZ_{G}$ is unknown. The problem can be formulated as
\begin{equation}\label{Prob:JointProbG}
\small
\begin{array}{lll}
\ds\max_{\bQ,\bA}\ds\frac{\mathbb{E}_{Z_{G}}\left[{\rm log}\left(\frac{\left|\sigma_{D}^{2}\bI_{N_{D}}+\bG\bA\left(\bH\bQ\bH^{H}+\sigma_{R}^{2}\bI_{N_{R}}\right)\bA^{H}\bG^{H}\right|}{\left|\sigma_{D}^{2}\bI_{N_{D}}+\sigma_{R}^{2}\bG\bA\bA^{H}\bG^{H}\right|}\right)\right]}{{\rm tr}\left(\bA(\bH\bQ\bH^{H}+\sigma_{R}^{2}\bI_{N_{R}})\bA^{H}\right)+{\rm tr}(\bQ)+P_{c}}
\\
{\rm s.t.\quad}{\rm tr}(\bA(\bH\bQ\bH^{H}+\sigma_{R}^{2}\bI_{N_{R}})\bA^{H})\leq P_{R}^{max}\\
{\rm tr}(\bQ)\leq P_{S}^{max} \;,\;\bQ\succeq 0\\
\mathbb{E}_{Z_{G}}\left[{\rm log}\left(\frac{\left|\sigma_{D}^{2}\bI_{N_{D}}+\bG\bA\left(\bH\bQ\bH^{H}+\sigma_{R}^{2}\bI_{N_{R}}\right)\bA^{H}\bG^{H}\right|}{\left|\sigma_{D}^{2}\bI_{N_{D}}+\sigma_{R}^{2}\bG\bA\bA^{H}\bG^{H}\right|}\right)\right]\geq R_{S}^{min}
\end{array}
\end{equation}
We will start by determining the optimal left and right eigenvector matrices of the AF matrix $\bA$ and the optimal transmit directions of the source covariance matrix $\bQ$.
\begin{proposition}
Consider Problem (\ref{Prob:JointProbG}). The optimal $\bQ$ and $\bA$ are such that $\bU_{Q}=\bV_{H}$, $\bU_{A}=\bU_{t,G}$ and $\bV_{A}=\bU_{H}$.
\end{proposition}
\begin{IEEEproof}
Exploiting the fact that the statistics of $\bZ_{G}$ do not change if $\bZ_{G}$ is multiplied, from left or right, by a unitary matrix, the numerator of the objective can be expressed as in (\ref{Eq:NumObjAG}).
\setcounter{equation}{29}
\begin{figure*}[!t]
\setcounter{MYtempeqncnt}{\value{equation}}
\setcounter{equation}{28}
\beq
\small
\begin{split}
N_{A,Q}&=\mathbb{E}_{Z_{G}}\left[{\rm log}\left|\sigma_{D}^{2}\bI_{N_{D}}+\bLambda_{r,G}^{1/2}\bZ_{G}\bLambda_{t,G}^{1/2}\bU_{t,G}^{H}\bA(\bH\bQ\bH^{H}+\sigma_{R}^{2}\bI_{N_{R}})\bA^{H}\bU_{t,G}\bLambda_{t,G}^{1/2}\bZ_{G}^{H}\bLambda_{r,G}^{1/2}\right|\right]\\
&-\mathbb{E}_{Z_{G}}\left[{\rm log}\left|\sigma_{D}^{2}\bI_{N_{D}}+\sigma_{R}^{2}\bLambda_{r,G}^{1/2}\bZ_{G}\bLambda_{t,G}^{1/2}\bU_{t,G}^{H}\bA\bA^{H}\bU_{t,G}\bLambda_{t,G}^{1/2}\bZ_{G}^{H}\bLambda_{r,G}^{1/2}\right|\right]\label{Eq:NumObjAG}
\end{split}
\eeq
\hrulefill
\setcounter{equation}{\value{MYtempeqncnt}}
\end{figure*}
Next, defining $\bY=\bH\bQ\bH^{H}$ and $\bX=\bLambda_{t,G}^{1/2}\bU_{t,G}^{H}\bA(\bY+\sigma_{R}^{2}\bI_{N_{R}})^{1/2}$, (\ref{Eq:NumObjAG}) can be rewritten as in (\ref{Eq:NumObjAG1}).
\setcounter{equation}{30}
\begin{figure*}[!t]
\setcounter{MYtempeqncnt}{\value{equation}}
\setcounter{equation}{29}
\beq
\small
\begin{split}
N_{X,Y}&=\mathbb{E}_{Z_{G}}\left[{\rm log}\left|\sigma_{D}^{2}\bI_{N_{D}}+\bLambda_{r,G}^{1/2}\bZ_{G}\bLambda_{x}\bZ_{G}^{H}\bLambda_{r,G}^{1/2}\right|\right]\\
&-\mathbb{E}_{Z_{G}}\left[{\rm log}\left|\sigma_{D}^{2}\bI_{N_{D}}+\sigma_{R}^{2}\bLambda_{r,G}^{1/2}\bZ_{G}\bLambda_{x}^{1/2}\bV_{x}^{H}(\bY+\sigma_{R}^{2}\bI_{N_{R}})^{-1}\bV_{x}\bLambda_{x}^{1/2}\bZ_{G}^{H}\bLambda_{r,G}^{1/2}\right|\right]\label{Eq:NumObjAG1}
\end{split}
\eeq
\hrulefill
\setcounter{equation}{\value{MYtempeqncnt}}
\end{figure*}
Let us define the two summands in (\ref{Eq:NumObjAG1}) as the functions $f_{1}$ and $f_{2}$. It is seen that $f_{1}$ does not depend on $\bU_{x}$ and $\bV_{x}$, while $f_{2}$ is a convex function of the matrix $\bM=\bV_{x}^{H}(\bY+\sigma_{R}^{2}\bI_{N_{R}})\bV_{x}$. Moreover, exploiting the fact that $\bGamma$ and $\bLambda_{x}^{1/2}$ commute because they are diagonal matrices of the same dimension and that $\bZ_{G}\bGamma$ has the same distribution as $\bZ_{G}$, it can be shown that $f_{2}(\bGamma\bM\bGamma)=f_{2}(\bM)$. Thus, by virtue of Lemma \ref{Lemma:DiagonalX} it follows that $f_{2}$ is minimized when $\bM$ is diagonal, which implies $\bV_{x}=\bU_{y}$. Thus, such choice maximizes (\ref{Eq:NumObjAG1}). 

As for the denominator of the objective of (\ref{Prob:JointProbG}), as a function of $\bX$ and $\bY$ it can be expressed as 
\beq\label{Eq:DenObjG}
\small
D_{X,Y}={\rm tr}(\bLambda_{t,G}^{-1/2}\bX\bX^{H}\bLambda_{t,G}^{-1/2})+{\rm tr}(\bH^{+}\bY\bH^{+H})+P_{c}\;.
\eeq
By virtue of Lemma \ref{Lemma:MinTrace}, (\ref{Eq:DenObjG}) is minimized when $\bU_{y}=\bU_{H}$, and $\bU_{x}=\bI_{N_{D}}$. Then we have $\bU_{x}=\bI_{N_{D}}$ and $\bV_{x}=\bU_{y}=\bU_{H}$. Moreover, these choices are also feasible, since they maximize the LHS of the QoS constraint and minimize the LHS of the power constraints.
Finally we obtain $\bY=\bU_{y}\bLambda_{y}\bU_{y}^{H}=\bU_{H}\bLambda_{H}^{1/2}\bV_{H}^{H}\bU_{Q}\bLambda_{Q}\bU_{Q}^{H}\bV_{H}\bLambda_{H}^{1/2}\bU_{H}^{H}$ and $\bX=\bU_{x}\bLambda_{x}\bV_{x}^{H}=\bLambda_{t,G}^{1/2}\bU_{t,G}^{H}\bU_{A}\bLambda_{A}^{1/2}\bV_{A}^{H}\bU_{y}(\bLambda_{y}+\sigma_{R}^{2}\bI_{N_{R}})^{1/2}\bU_{y}^{H}$, from which it follows that the conditions $\bU_{x}=\bI_{N_{D}}$ and $\bV_{x}=\bU_{y}=\bU_{H}$ are fulfilled if $\bU_{Q}=\bV_{H}$, $\bU_{A}=\bU_{t,G}$, and $\bV_{A}=\bU_{H}$. 
\end{IEEEproof}
Thus, we are left with the power control optimization problem shown in (\ref{Prob:JointProbFinalG}),
\setcounter{equation}{32}
\begin{figure*}[!t]
\setcounter{MYtempeqncnt}{\value{equation}}
\setcounter{equation}{31}
\begin{equation}\label{Prob:JointProbFinalG}
\small
\left\{
\begin{array}{lll}
\ds\max_{\bLambda_{Q},\bLambda_{A}}\frac{\ds\mathbb{E}_{Z_{G}}\left[{\rm log}\left(\frac{\left|\sigma_{D}^{2}\bI_{N_{D}}+\bLambda_{r,G}\bZ_{G}\bLambda_{t,G}\bLambda_{A}\left(\bLambda_{H}^{1/2}\bLambda_{Q}\bLambda_{H}^{H/2}+\sigma_{R}^{2}\bI_{N_{R}}\right)\bZ_{G}^{H}\right|}{\left|\sigma_{D}^{2}\bI_{N_{D}}+\sigma_{R}^{2}\bLambda_{r,G}\bZ_{G}\bLambda_{t,G}\bLambda_{A}\bZ_{G}^{H}\right|}\right)\right]}{{\rm tr}\left(\bLambda_{A}\left(\bLambda_{H}^{1/2}\bLambda_{Q}\bLambda_{H}^{H/2}+\sigma_{R}^{2}\bI_{N_{R}}\right)\right)+{\rm tr}(\bLambda_{Q})+P_{c}}
\\
{\rm s.t.\quad}{\rm tr}\left(\bLambda_{A}\left(\bLambda_{H}^{1/2}\bLambda_{Q}\bLambda_{H}^{H/2}+\sigma_{R}^{2}\bI_{N_{R}}\right)\right)\leq P_{R}^{max}\;,\;\;
{\rm tr}(\bLambda_{Q})\leq P_{S}^{max} \;,\;\bLambda_{Q}\succeq 0\;,\;\bLambda_{A}\succeq 0 \\
\ds\mathbb{E}_{Z_{G}}\left[{\rm log}\left(\frac{\left|\sigma_{D}^{2}\bI_{N_{D}}+\bLambda_{r,G}\bZ_{G}\bLambda_{t,G}\bLambda_{A}\left(\bLambda_{H}^{1/2}\bLambda_{Q}\bLambda_{H}^{H/2}+\sigma_{R}^{2}\bI_{N_{R}}\right)\bZ_{G}^{H}\right|}{\left|\sigma_{D}^{2}\bI_{N_{D}}+\sigma_{R}^{2}\bLambda_{r,G}\bZ_{G}\bLambda_{t,G}\bLambda_{A}\bZ_{G}^{H}\right|}\right)\right]\geq R_{S}^{min}
\end{array}
\right.
\end{equation}
\hrulefill
\setcounter{equation}{\value{MYtempeqncnt}}
\end{figure*}
which is neither jointly concave nor jointly pseudo-concave with respect to both $\bLambda_{Q}$ and $\bLambda_{A}$. However, it can be shown to be separately pseudo-concave in $\bLambda_{Q}$ and $\bLambda_{A}$. Pseudo-concavity with respect to $\bLambda_{Q}$ is clear, since for any fixed $\bLambda_{A}$, the numerator of the objective is concave and the denominator is linear in $\bLambda_{Q}$, while the constraints are all linear or concave. As for the pseudo-concavity with respect to $\bLambda_{A}$, it is clear that the denominator is linear in $\bLambda_{A}$. Then, we need to show the concavity of the numerator. To this end, we rewrite the numerator as $\small\mathbb{E}_{F}\left[{\rm log}\left|\bI_{N_{D}}+(\sigma_{D}^{2}\bI_{N_{D}}+\sigma_{R}^{2}\widetilde{\bF}\bLambda_{A}\widetilde{\bF}^{H})^{-1}\widetilde{\bF}\bLambda_{A}\bLambda_{B}\widetilde{\bF}^{H}\right|\right]$, with $\widetilde{\bF}=\bLambda_{r,G}^{1/2}\bZ_{G}\bLambda_{t,G}^{1/2}$ and $\bLambda_{B}=\bLambda_{H}^{1/2}\bLambda_{Q}\bLambda_{H}^{H/2}$, which is concave in $\bLambda_{A}$ as shown in \cite{ZapRelayMIMO2013}. Consequently, an alternating maximization algorithm to solve Problem (\ref{Prob:JointProbFinalG}) can be devised in a similar fashion as Algorithm \ref{Alg:AltMaxSCSI} as follows.
\begin{algorithm}
\small
\caption{Alternating maximization for Problem (\ref{Prob:JointProbFinalG})}
\label{Alg:GAltMaxSCSI}
\begin{algorithmic}
\STATE
Initialize $\bLambda_{Q}^{(0)}$ to a feasible value. Set a tolerance $\epsilon$. Set $n=0$;
\REPEAT
\STATE Given $\bLambda_{Q}^{(n)}$, solve Problem (\ref{Prob:JointProbFinalG}) with respect to $\bLambda_{A}$ to obtain the optimal $\bLambda_{A}^{(n+1)}$;
\STATE Given $\bLambda_{A}^{(n+1)}$, solve Problem (\ref{Prob:JointProbFinalG}) with respect to $\bLambda_{Q}$ to obtain the optimal $\bLambda_{Q}^{(n+1)}$;
\STATE $n=n+1$;
\UNTIL{$\left|{\rm GEE}^{(n)}-{\rm GEE}^{(n)}\right| \leq \epsilon$}
\end{algorithmic}
\end{algorithm}

\section{Optimality of full-power beamforming.}\label{Sec:OptBeam}
In this section, for both the scenarios of Sections \ref{Sec:Uplink} and \ref{Sec:Downlink}, conditions under which the optimal source power allocation policy is to support only one data stream are determined. Otherwise stated, conditions for the solution of Problems (\ref{Prob:JointProbH2}) and (\ref{Prob:JointProbFinalG}), with respect to $\bLambda_{Q}$, to be a unit-rank matrix will be derived.
Conditions for single-stream transmission optimality are usually referred to as beamforming optimality, and have been previously investigated with reference to achievable rate maximization problems \cite{jafar2004transmitter,jorswieck2004channel,soysal2009optimality,DharmawansaBeam2011,AhamadiBeam2011,BeamMISO} motivated by the considerations that using just one transmission stream allows to make use of the well-developed theory of channel coding for SISO systems, simplifies the receiver design since just one data stream needs to be decoded, and, above all, it greatly simplifies the resource allocation process. Indeed, in rate-maximization problems, beamforming optimality implies that all of the available power should be concentrated on the strongest channel eigenvalue, and therefore the optimal source covariance matrix is immediately determined in closed-form, without having to solve any optimization problem. Unfortunately, this is not entirely true when dealing with GEE maximization, because even in a scenario in which supporting only one data stream is optimal, the GEE might not be maximized for full-power transmission. Therefore, as far as GEE maximization is concerned, two different beamforming optimality problems arise: 1) full-power (FP) beamforming optimality, in which conditions are determined such that the solution of Problems (\ref{Prob:JointProbH2}) and (\ref{Prob:JointProbFinalG}) is not only to support one data stream, but also to concentrate all of the available power on such data stream, and 2) non full-power (NFP) beamforming optimality, in which the solution of Problems (\ref{Prob:JointProbH2}) and (\ref{Prob:JointProbFinalG}) is only required to be a generic rank-one matrix. In the following, the focus will be on the former problem, which provides conditions that, if fulfilled, immediately allow to determine the optimal source power allocation vector as $\blambda_{Q}=(P,0,\ldots,0)$, with $P$ being the maximum feasible power for the first component of $\blambda_{Q}$, without having to actually solve any source optimization problem. Instead, deriving conditions for NFP beamforming only allows to characterize the optimal $\blambda_{Q}$ as $\blambda_{Q}=(p,0,\ldots,0)$, with $p\leq P$ being the optimal power level to be still determined by solving the GEE maximization problem. It should also be stressed that since FP beamforming optimality is a special case of NFP beamforming, the conditions to be derived for FP beamforming optimality will also be sufficient conditions for NFP beamforming optimality. 

\subsection{FP beamforming with partial CSI on $\bH$}\label{Sec:BeamUp}
Consider Problem (\ref{Prob:JointProbH2}) with respect to $\bLambda_{Q}$. First of all, in order to obtain mathematically tractable 
conditions, the QoS constraint will be relaxed as far as deriving beamforming conditions is concerned. It should be observed that the resulting problem is an EE maximization problem subject to power constraints, a problem which has been tackled in many previous papers with reference to different communication systems (see for example \cite{Betz2008},\cite{Meshkati2007},\cite{Miao2011},\cite{belmega2011energy},\cite{MIMOBroadcastEE},\cite{ZapponeTWC}). Moreover, finding conditions for the relaxed problem might prove useful also with reference to the original problem. Indeed, if source beamforming is optimal for the relaxed problem, one can simply check wether such solution fulfills the QoS constraint. If yes, then source beamforming will also be optimal for the original problem. 

Now, defining $\bF=\bLambda_{C}^{1/2}\bZ_{H}$, with $\bLambda_{C}=\widetilde{\bLambda}_{y}\left(\sigma_{D}^{2}\bI_{N_{R}}+\sigma_{R}^{2}\widetilde{\bLambda}_{y}\bLambda_{r,H}^{-1}\right)^{-1}$ and $\widetilde{\bLambda}_{y}$ defined as in Section \ref{Sec:Uplink}, $b=P_{c}+\sigma_{R}^{2}{\rm tr}\left(\widetilde{\bLambda}_{y}\bLambda_{r,H}^{-1}\widetilde{\bLambda}_{G}\right)$, $c={\rm tr}\left(\widetilde{\bLambda}_{y}\widetilde{\bLambda}_{G}\right)$, and applying the change of variables $\lambda_{i}=\lambda_{i,Q}\lambda_{i,H}^{t}$, for all $i=1,\ldots,N_{S}$, (\ref{Prob:JointProbH2}) can be equivalently recast as
\beq\label{Prob:OptCovHJoint2}
\left\{
\begin{array}{lll}
\ds\max_{\{\lambda_{i}\}_{i=1}^{N_{S}}}\frac{\mathbb{E}_{\{\bbf_{i}\}_{i=1}^{N_{S}}}\left[{\rm log}\left|\bI_{N_{R}}+\sum_{i=1}^{N_{S}}\lambda_{i}\bbf_{i}\bbf_{i}^{H}\right|\right]}{b+\sum_{i=1}^{N_{S}}\lambda_{i}d_{i}}
\\
{\rm s.t.\quad}\sum_{i=1}^{N_{S}}\lambda_{i}\leq \ds\frac{P_{R}^{max}+P_{c}-b}{c}\\
\sum_{i=1}^{N_{S}}\frac{\lambda_{i}}{\lambda_{i,H}^{t}}\leq P_{S}^{max}\;,\;\;\lambda_{i}\geq 0 \;,\;\forall\;i=1,\ldots,N_{S}\\
\end{array}
\right. \;,
\eeq
%\beq\label{Prob:OptCovHJoint1}
%\small
%\left\{
%\begin{array}{lll}
%\ds\max_{\blambda_{Q}}\frac{\mathbb{E}_{\{\bbf_{i}\}_{i=1}^{N_{S}}}\left[{\rm log}\left|\bI_{N_{R}}+\sum_{i=1}^{N_{S}}\lambda_{i,Q}\lambda_{i,H}^{t}\bbf_{i}\bbf_{i}^{H}\right|\right]}{b+\sum_{i=1}^{N_{S}}\lambda_{i,Q}(c\lambda_{i,H}^{t}+1)}
%\\
%{\rm s.t.\quad}\sum_{i=1}^{N_{S}}\lambda_{i,Q}\lambda_{i,H}^{t}\leq \ds\frac{P_{R}^{max}+P_{c}-b}{c}\\
%\sum_{i=1}^{N_{S}}\lambda_{i,Q}\leq P_{S}^{max}\;,\;\;\lambda_{i,Q}\geq 0 \;,\;\forall\;i=1,\ldots,N_{S}\\
%\end{array}
%\right. \;,
%\eeq
wherein $\{\bbf_{i}\}_{i=1}^{N_{S}}$ are the first $N_{S}$ columns of $\bF$, which are therefore i.i.d. Gaussian vectors with covariance matrix $\mathbb{E}_{\bbf_{i}}\left[\bbf_{i}\bbf_{i}^{H}\right]=\bLambda_{C}$ for all $i=1,\ldots,N_{S}$, and $d_{i}=c+\ds\frac{1}{\lambda_{i,H}^{t}}$. Then, the following proposition holds.
\begin{proposition}\label{Prop:BeamOpt}
For all $i=1,\ldots,N_{S}$, define
\beq
\small
\begin{split}
C_{i,H}=&\left({\rm tr}(\bLambda_{C})-P\mathbb{E}_{\bbf_{1}}\left[\frac{\bbf_{1}^{H}\bLambda_{C}\bbf_{1}}{1+P\|\bbf_{1}\|^{2}}\right]\right)(b+Pd_{1})\\
&-d_{i}\mathbb{E}_{\bbf_{1}}\left[{\rm log}\left(1+P\|\bbf_{1}\|^{2}\right)\right]\;.
\end{split}
\eeq
Define also $P=\min\left(P_{S}^{max}\lambda_{1,H}^{t},\ds\frac{P_{R}^{max}+P_{c}-b}{c}\right)$.
Then, if $C_{2,H}\geq 0$, then FP beamforming is optimal if and only if 
\beq\label{Eq:CondBeam1}
\small
\begin{split}
&P\left({\rm tr}(\bLambda_{C})-P\mathbb{E}_{\bbf_{1}}\left[\frac{\bbf_{1}^{H}\bLambda_{C}\bbf_{1}}{1+P\|\bbf_{1}\|^{2}}\right]\right)+\mathbb{E}_{\bbf_{1}}\left[\frac{1}{1+P\|\bbf_{1}\|^{2}}\right]\\
&+\frac{P(d_{1}-d_{2})}{b+Pd_{1}}\mathbb{E}_{\bbf_{1}}\left[{\rm log}\left(1+P\|\bbf_{1}\|^{2}\right)\right]\leq 1
\end{split}
\eeq
Instead, if $C_{2,H} < 0$, then FP beamforming is optimal if and only if
\beq\label{Eq:CondBeam2}
\small
\frac{Pd_{1}}{b+Pd_{1}}\mathbb{E}_{\bbf_{1}}\left[{\rm log}\left(1+P\|\bbf_{1}\|^{2}\right)\right]+\mathbb{E}_{\bbf_{1}}\left[\frac{1}{1+P\|\bbf_{1}\|^{2}}\right]\leq 1\;.
\eeq
\end{proposition}
\begin{IEEEproof}
To begin with, let us observe that the two power constraints of (\ref{Prob:OptCovHJoint2}) define two $N_{S}$-dimensional hyperplanes whose $N_{S}$ intersection points with the $N_{S}$ axis are given by $\ds\frac{P_{R}^{max}+P_{c}-b}{c}$, for all $i=1,\ldots,N_{S}$, and $\{P_{S}^{max}\lambda_{i,H}^{t}\}_{i=1}^{N_{S}}$, respectively. Consequently, for all $i=1,\ldots,N_{S}$, the power available in the direction associated to $\lambda_{i}$ is $\ds\min\left(P_{S}^{max}\lambda_{i,H}^{t},\frac{P_{R}^{max}+P_{c}-b}{c}\right)$. Thus, since $\lambda_{1,H}^{t}\geq\lambda_{i,H}^{t}$ and $d_{1}\leq d_{i}$ for all $i=1,\ldots,N_{S}$, it is seen that the optimal transmit direction is $\bbf_{1}$, and that the power available in this direction is $P$. Then, let us consider the power allocation policy $\blambda=(P-p,\alpha_{2}p,\ldots,\alpha_{N_{S}}p)$, with $\sum_{i=2}^{N_{S}}\alpha_{i}\leq 1$. Rewriting the objective function accordingly yields
\beq
\small
g(p)=\frac{\mathbb{E}_{F}\left[{\rm log}\left|\bI_{N_{R}}+P\bbf_{1}\bbf_{1}^{H}+p\left(\sum_{i=2}^{N_{S}}\alpha_{i}\bbf_{i}\bbf_{i}-\bbf_{1}\bbf_{1}^{H}\right)\right|\right]}{b+Pd_{1}+p\left(\sum_{i=2}^{N_{S}}\alpha_{i}d_{i}-d_{1}\right)}\;.
\eeq
FP beamforming is optimal if and only if the function $g(p)$ is maximized at $p=0$, and, since $g(p)$ is a pseudo-concave function for $p\geq 0$, a necessary and sufficient condition for this to happen is $\frac{\partial g}{\partial p}_{|p=0}\leq 0$ for all $\{\alpha_{i}\}_{i=2}^{N_{S}}$ such that $\sum_{i=2}^{N_{S}}\alpha_{i}\leq 1$. Then, exploiting Lemma \ref{LemmaDerLogDet} and elaborating, we obtain the condition
\begin{gather}
\small
\mathbb{E}_{F}\left[{\rm tr}\left(\left(\bI_{N_{R}}+P\bbf_{1}\bbf_{1}^{H}\right)^{-1}\left(\sum_{i=2}^{N_{S}}\alpha_{i}\bbf_{i}\bbf_{i}-\bbf_{1}\bbf_{1}^{H}\right)\right)\right]\times\notag\\
\small
\label{Eq:CondDerBeam}
\times(b+Pd_{1})+\left(d_{1}-\sum_{i=2}^{N_{S}}\alpha_{i}d_{i}\right)\mathbb{E}_{\bbf_{1}}\left[{\rm log}\left(1+P\|\bbf_{1}\|^{2}\right)\right]\leq 0\;.
\end{gather}
Focusing on the first statistical average in (\ref{Eq:CondDerBeam}) we have
\begin{gather}
\small
\mathbb{E}_{F}\left[{\rm tr}\left(\left(\bI_{N_{R}}+P\bbf_{1}\bbf_{1}^{H}\right)^{-1}\sum_{i=2}^{N_{S}}\alpha_{i}\bbf_{i}\bbf_{i}\right)\right]\notag\\
\small
-\mathbb{E}_{F}\left[{\rm tr}\left(\left(\bI_{N_{R}}+P\bbf_{1}\bbf_{1}^{H}\right)^{-1}(P\bbf_{1}\bbf_{1}^{H}+\bI_{N_{R}}-\bI_{N_{R}})\frac{1}{P}\right)\right]\notag\\
\small
=\sum_{i=2}^{N_{S}}\alpha_{i}\mathbb{E}_{\bbf_{1}}\left[{\rm tr}\left(\left(\bI_{N_{R}}+P\bbf_{1}\bbf_{1}^{H}\right)^{-1}\bLambda_{C}\right)\right]-\frac{N_{R}}{P}\notag\\
\small
+\frac{1}{P}\mathbb{E}_{\bbf_{1}}\left[{\rm tr}\left(\left(\bI_{N_{R}}+P\bbf_{1}\bbf_{1}^{H}\right)^{-1}\right)\right]\;,
\end{gather}
where we have exploited the fact that $\{\bbf_{i}\}_{i=1}^{N_{S}}$ are statistically i.i.d. random vectors, with covariance matrix $\bLambda_{C}$. Then, applying the matrix inversion lemma and rearranging terms, (\ref{Eq:CondDerBeam}) can be rewritten as
\beq
\small
\begin{split}
\label{Eq:CondDerBeam2}
\sum_{i=2}^{N_{S}}\alpha_{i}C_{i,H}\leq &\frac{(b+Pd_{1})}{P}\left(1-\mathbb{E}_{\bbf_{1}}\left[\frac{1}{1+P\|\bbf_{1}\|^{2}}\right]\right)\\
&-d_{1}\mathbb{E}_{\bbf_{1}}\left[{\rm log}\left(1+P\|\bbf_{1}\|^{2}\right)\right]
\end{split}
\eeq
Now, since (\ref{Eq:CondDerBeam2}) needs to hold for all $\{\alpha_{i}\}_{i=2}^{N_{S}}$ such that $\sum_{i=2}^{N_{S}}\alpha_{i}\leq 1$, the worst case in which the LHS of (\ref{Eq:CondDerBeam2}) is maximized with respect to $\{\alpha_{i}\}_{i=2}^{N_{S}}$ needs to be considered. Since $C_{2,H}\geq C_{3,H}\geq\ldots\geq C_{N_{S},H}$, if $C_{2,H}\geq 0$, then the LHS of (\ref{Eq:CondDerBeam2}) is maximized with respect to $\{\alpha_{i}\}_{i=2}^{N_{S}}$ when $\alpha_{2}=1$ and $\alpha_{i}=0$ for all $i=3,\ldots,N_{S}$, which yields (\ref{Eq:CondBeam1}). Instead, if $C_{2,H}$ is negative, then the LHS of (\ref{Eq:CondDerBeam2}) is maximized by setting $\alpha_{i}=0$, for all $i=2,\ldots,N_{S}$, which yields (\ref{Eq:CondBeam2}).
\end{IEEEproof}
Proposition \ref{Prop:BeamOpt} provides a condition that allows to check wether single-stream transmission is the optimal power allocation strategy for Problem (\ref{Prob:OptCovHJoint2}). The condition can be checked before starting to solve Problem (\ref{Prob:OptCovHJoint2}) and, if fulfilled, we immediately obtain the solution to Problem (\ref{Prob:OptCovHJoint2}) without actually having to solve it. Moreover, we can exploit all the advantages of single-stream transmission, as detailed at the beginning of Section \ref{Sec:OptBeam}, without losing from an energy-efficient point of view.

\subsection{FP beamforming with partial CSI on $\bG$}\label{Sec:BeamDown}
Consider Problem (\ref{Prob:JointProbFinalG}) with respect to $\bLambda_{Q}$. Similarly to Section \ref{Sec:BeamUp}, the QoS constraint will be relaxed. Moreover, we assume that $\bLambda_{A}$ is chosen so as to equalize $\bLambda_{t,G}$. Otherwise stated we assume $\bLambda_{t,G}\bLambda_{A}=\bI_{N_{R}}$. Then, defining $\bF=\left(\sigma_{D}^{2}\bI_{N_{D}}+\sigma_{R}^{2}\bLambda_{r,G}^{1/2}\bZ_{G}\bZ_{G}^{H}\bLambda_{r,G}^{1/2}\right)^{-1/2}\bLambda_{r,G}^{1/2}\bZ_{G}$, $b=\sigma_{R}^{2}\sum_{i=1}^{N_{R}}\lambda_{i,A}+P_{c}$, $d_{i}=1+\frac{1}{\lambda_{i,A}\lambda_{i,H}}$, and applying the change of variables $\lambda_{i}=\lambda_{i,Q}\lambda_{i,H}\lambda_{i,A}$, for all $i=1,\ldots,N_{S}$, Problem (\ref{Prob:JointProbFinalG}) can be restated as 
\beq\label{Prob:OptCovGJoint2}
\small
\left\{
\begin{array}{lll}
\ds\max_{\{\lambda_{i}\}_{i=1}^{N_{S}}}\frac{\mathbb{E}_{\{\bbf_{i}\}_{i=1}^{N_{S}}}\left[{\rm log}\left|\bI_{N_{D}}+\sum_{i=1}^{N_{S}}\lambda_{i,G}^{t}\lambda_{i}\bbf_{i}\bbf_{i}^{H}\right|\right]}{b+\sum_{i=1}^{N_{S}}\lambda_{i}d_{i}}
\\
{\rm s.t.\quad} \sum_{i=1}^{N_{S}}\lambda_{i}\leq P_{R}^{max}+P_{c}-b
\\
\sum_{i=1}^{N_{S}}\frac{\lambda_{i}}{\lambda_{i,H}\lambda_{i,A}}\leq P_{S}^{max}\;, \quad\lambda_{i}\geq 0\;, \forall i=1,\ldots,N_{S}
\end{array}
\right. \;.
\eeq
%\beq\label{Prob:OptCovGJoint1}
%\small
%\left\{
%\begin{array}{lll}
%\ds\max_{\blambda_{Q}}\frac{\mathbb{E}_{\{\bbf_{i}\}_{i=1}^{N_{S}}}\left[{\rm log}\left|\bI_{N_{D}}+\sum_{i=1}^{N_{S}}\lambda_{i,G}^{t}\lambda_{i,A}\lambda_{i,H}\lambda_{i,Q}\bbf_{i}\bbf_{i}^{H}\right|\right]}{b+\sum_{i=1}^{N_{S}}\lambda_{i,Q}(1+\lambda_{i,A}\lambda_{i,H})}
%\\
%{\rm s.t.\quad} \sum_{i=1}^{N_{S}}\lambda_{i,A}\lambda_{i,H}\lambda_{i,Q}\leq P_{R}^{max}+P_{c}-b
%\\
%\sum_{i=1}^{N_{S}}\lambda_{i,Q}\leq P_{S}^{max}\;, \quad\lambda_{i,Q}\geq 0\;, \forall i=1,\ldots,N_{S}
%\end{array}
%\right. \;.
%\eeq
Then, the following proposition holds.
\begin{proposition}
For all $i=1,\ldots,N_{S}$, define 
\begin{gather}
\small
C_{i,G}=\lambda_{i,G}^{t}\left(\mathbb{E}_{\bbf_{2}}\left[\|\bbf_{2}\|^{2}\right]-P\lambda_{1,G}^{t}\mathbb{E}_{\bbf_{1},\bbf_{2}}\left[\frac{|\bbf_{2}^{H}\bbf_{1}|^{2}}{1+P\lambda_{1,G}^{t}\|\bbf_{1}\|^{2}}\right]\right)\notag\\
\times(b+Pd_{1})-d_{i}\mathbb{E}_{\bbf_{1}}\left[{\rm log}\left(1+P\lambda_{1,G}^{t}\|\bbf_{1}\|^{2}\right)\right]\;.
\end{gather}
Define also $P=\min\left(\lambda_{1,A}\lambda_{1,H}P_{S}^{max},P_{R}^{max}+P_{c}-b\right)$. If $C_{2,G}\geq 0$, then FP beamforming is optimal if and only if 
\begin{gather}\label{Eq:CondBeamG1}
\small
P\lambda_{2,G}^{t}\left(\mathbb{E}_{\bbf_{2}}\left[\|\bbf_{2}\|^{2}\right]-P\lambda_{1,G}^{t}\mathbb{E}_{\bbf_{1},\bbf_{2}}\left[\frac{|\bbf_{2}^{H}\bbf_{1}|^{2}}{1+P\lambda_{1,G}^{t}\|\bbf_{1}\|^{2}}\right]\right)\notag\\
\small
+\mathbb{E}_{\bbf_{1}}\left[\frac{1}{1+P\lambda_{1,G}^{t}\|\bbf_{1}\|^{2}}\right]\notag\\
\small
+\frac{P(d_{1}-d_{2})}{b+Pd_{1}}\mathbb{E}_{\bbf_{1}}\left[{\rm log}\left(1+P\lambda_{1,G}^{t}\|\bbf_{1}\|^{2}\right)\right]\leq 1
\end{gather}
Instead, if $C_{2,G}\leq0$, then FP beamforming is optimal if and only if 
\begin{gather}
\label{Eq:CondBeamG2}
\small
\mathbb{E}_{\bbf_{1}}\left[\frac{1}{1+P\lambda_{1,G}^{t}\|\bbf_{1}\|^{2}}\right]\\
\small
+\frac{Pd_{1}}{b+Pd_{1}}\mathbb{E}_{\bbf_{1}}\left[{\rm log}\left(1+P\lambda_{1,G}^{t}\|\bbf_{1}\|^{2}\right)\right]\leq 1
\;.
\end{gather}
\end{proposition}
\begin{IEEEproof}
The proof follows similar arguments as the proof of Proposition \ref{Prop:BeamOpt}. The main difference is that now the columns of the matrix $\bF$ are not independent with one another. However, they are still identically distributed, which can be exploited as shown in the following. 
First of all, it is seen that the optimal transmit direction is $\bbf_{1}$ and the available power in this direction is $P$. Then, considering the power allocation policy $\blambda={\rm diag}(P-p,\alpha_{2}p,\ldots,\alpha_{N_{S}}p)$, with $\sum_{i=2}^{N_{S}}\alpha_{i}\leq 1$ and evaluating the condition for the first derivative of the objective function to be non-positive in $p=0$, yields
\beq
\small
\begin{split}
&\Bigg\{\mathbb{E}_{\{\bbf_{i}\}_{i=1}^{N_{S}}}\left[{\rm tr}\left(\left(\bI_{N_{D}}+P\lambda_{1,G}^{t}\bbf_{1}\bbf_{1}^{H}\right)^{-1}\sum_{i=2}^{N_{S}}\alpha_{i}\lambda_{i,G}^{t}\bbf_{i}\bbf_{i}\right)\right]\\
&-\frac{N_{D}}{P}+\frac{1}{P}\mathbb{E}_{\bbf_{1}}\left[{\rm tr}\left(\left(\bI_{N_{D}}+P\lambda_{1,G}^{t}\bbf_{1}\bbf_{1}^{H}\right)^{-1}\right)\right]\Bigg\}(b+Pd_{1})\\
&+\left(d_{1}-\sum_{i=2}^{N_{S}}\alpha_{i}d_{i}\right)\mathbb{E}_{\bbf_{1}}\left[{\rm log}\left(1+P\lambda_{1,G}^{t}\|\bbf_{1}\|^{2}\right)\right]\leq 0\;.
\label{Eq:CondDerBeamG}
\end{split}
\eeq
Now, applying the matrix inversion lemma and exploiting the fact that the columns of $\bF$ are identically distributed, (\ref{Eq:CondDerBeamG}) can be rewritten as
\begin{gather}
\label{Eq:CondDerBeamG1}
\small
\sum_{i=2}^{N_{S}}\alpha_{i}C_{i,G}\leq \frac{(b+Pd_{1})}{P}\left(1-\mathbb{E}_{\bbf_{1}}\left[\frac{1}{1+P\lambda_{1,G}^{t}\|\bbf_{1}\|^{2}}\right]\right)-\notag\\
\small
-d_{1}\mathbb{E}_{\bbf_{1}}\left[{\rm log}\left(1+P\lambda_{1,G}^{t}\|\bbf_{1}\|^{2}\right)\right]\;,
\end{gather}
and the thesis follows by the same argument of Proposition \ref{Prop:BeamOpt}.
\end{IEEEproof}

\section{Numerical results}\label{Sec:NumericalResults}
In our simulations a relay-assisted system with $N_{S}=N_{R}=N_{D}=3$ antennas has been considered. It has been set $\sigma_{R}^{2}=\sigma_{D}^{2}=\sigma^{2}$, $P_{S}^{max}=P_{R}^{max}=P^{max}$, and the performance has been evaluated in terms of the achieved instantaneous GEE given by (\ref{Eq:GEE}), versus the SNR defined as ${\rm SNR}=\frac{P^{max}}{\sigma^{2}}$. The QoS constraint has been set to $R_{S}^{min}=1$ bit/s/Hz and the total circuit power to $P_{c}=5$W. The transmit and receive correlation matrices $\bR_{r,H}$, $\bR_{t,H}$, $\bR_{r,G}$, and $\bR_{t,G}$ have been generated according to the exponential correlation model, with equal correlation index $\rho$, whereas the matrices $\bZ_{H}$ and $\bZ_{G}$ have been generated as realizations of Gaussian matrices with zero-mean and unit-variance proper complex Gaussian entries. 

In Figs. \ref{fig:1} and \ref{fig:2} the required number of iterations needed for the proposed alternating maximization algorithms to converge is shown for the two scenarios of statistical CSI on $\bH$ and $\bG$, respectively, for ${\rm SNR}=20$dB, $\rho=0.5$, and a tolerance $\epsilon=10^{-3}$. For each scenario, the algorithms have been run from a set of $10$ different, randomly selected, initialization points and the instantaneous GEE value achieved in each iteration for each initialization point is reported.
First of all, it is seen that for both the considered CSI assumptions, convergence occurs after a few iterations. Secondly, it is seen that regardless of the initialization point, the proposed algorithms converge to a unique fixed point\footnote{As far as algorithm \ref{Alg:AltMaxSCSI} is concerned, a similar behavior has been also observed when increasing the number of initialization points. Instead, as for Algorithm \ref{Alg:GAltMaxSCSI}, it has been observed that there is a small probability that it converges to a different fixed point.}, which confirms the high SNR considerations made in Section \ref{Sec:PerfectCSI}. Instead, in the low SNR regime the situation is different and the initialization point influences the fixed point of the algorithm. For this reason, in the following illustrations, the performance of the proposed algorithms has been always evaluated by considering $10$ randomly selected initialization points and then picking the initialization point resulting in the best fixed point. Moreover, the presented results have been obtained by averaging over $1000$ independent scenarios.

In Fig. \ref{fig:3} the achieved instantaneous GEE($\bQ,\bA$) versus the ${\rm SNR}$ is illustrated for $\rho=0.5$, and for $(\bQ,\bA)$ obtained from the following resource allocation algorithms.
\begin{enumerate}
\item
$\bQ$ and $\bA$ resulting from resource allocation with perfect CSI on both $\bH$ and $\bG$.
\item $\bQ$ and $\bA$ resulting from resource allocation with perfect CSI on $\bG$ and statistical CSI on $\bH$, namely Algorithm \ref{Alg:AltMaxSCSI}. 
\item $\bQ$ and $\bA$ resulting from resource allocation with perfect CSI on $\bH$ and statistical CSI on $\bG$, namely Algorithm \ref{Alg:GAltMaxSCSI}.
\end{enumerate}
As expected, better performance is obtained when perfect CSI is available. Interestingly, it is also seen that having perfect knowledge of $\bH$ grants better performance than when $\bG$ is perfectly available. This can be explained noticing that the source-to-relay channel $\bH$ affects both the numerator and the denominator of the GEE, whereas the relay-to-destination channel $\bG$ affects only the numerator.

In Fig. \ref{fig:4} a similar scenario as in Fig. \ref{fig:3} has been considered, with the difference that the performance achieved for $\rho=0.1$ and $\rho=0.9$ is contrasted. As for $\rho=0.5$, having perfect knowledge of $\bH$ allows to achieve a higher instantaneuos GEE than when $\bG$ is perfectly available. Moreover, the results indicate that the gap to the perfect CSI case is bigger when $\rho=0.1$ than when $\rho=0.9$. This is also expected because the correlation index $\rho$ is a measure of how much information we have on the channel.

Finally, in Fig. \ref{fig:5} the beamforming condition derived in Section \ref{Sec:BeamUp} with statistical CSI on $\bH$ is validated. Fig. \ref{fig:4} considers a system with $N_{S}=N_{R}=N_{D}=2$, $P_{S}^{max}=1$ Watt, $\bLambda_{C}=\bI_{N_{R}}$, $\lambda_{t,1}=2$, $\lambda_{t,2}=1$, $b=0.1$, and $c=0.5$. With these parameters, Problem (\ref{Prob:OptCovHJoint2}) has been solved for increasing values of $P_{R}^{max}$, and the components $\lambda_{1}$ and $\lambda_{2}$ of the solution, normalized so as to add up to $1$, have been plotted versus the parameter $P$, defined in the proof of Proposition \ref{Prop:BeamOpt}. The black line in the plot marks the beamforming optimality region computed according to the condition in Proposition \ref{Prop:BeamOpt}, and indeed it is seen that when such limit is exceeded, some power starts being allocated to the second transmission stream, too.

\section{Conclusions}\label{Sec:Conclusion}
EE optimization in relay-assisted systems has been carried out in this paper assuming both perfect and statistical CSI. The considered energy-efficient performance function is the system GEE, defined as the ratio between the achievable rate and the consumed power. 
First, the optimal source and relay eigenvector matrices have been determined and it has been shown that diagonalization of both the numerator and the denominator of the GEE is optimal. Next, it has been shown that the resulting power allocation problem is separately pseudo-concave in the source and relay power allocation vector, and therefore has been tackled by means of fractional programming and alternating maximization. With reference to the statistical CSI scenarios, conditions for beamforming optimality have also been derived. The performance of all of the considered scenarios have been numerically addressed and contrasted. 

It should be remarked that also the more general case in which both the source-relay and the relay-destination channels are only statistically known could be tackled by means of the same techniques developed in this paper. Details are omitted due to space constraints, but it is possible to show that diagonalizing the correlation matrices of the statistically available channels is optimal in this scenario, too. However, the resulting power control problem can not be shown to be separately pseudo-concave in the source and relay power vectors. Thus, such a scenario deserves further investigation as far as the design of low-complexity power control algorithms is concerned. 

\appendix
\begin{lemma}\label{Lemma:MinTrace}
The function ${\rm tr}(\bT\bR)$, where $\bT$ and $\bR$ are Hermitian matrices with proper dimensions, is minimized when $\bT$ and $\bR$ commute and have eigenvalues arranged in opposite order.
\end{lemma}
\begin{IEEEproof}
See \cite[Lemma H.1.h]{marshall1979inequalities}.
\end{IEEEproof}

\begin{lemma}\label{Lemma:MinDet}
The function ${\rm log}|\bI_{N}+\bT^{-1}\bR|$, where $\bT$ is $N \times N$, Hermitian, and positive definite, while $\bR$ is $N \times N$, Hermitian, and positive semidefinite, is minimized when $\bT$ and $\bR$ commute and have eigenvalues arranged in the same order.
\end{lemma}
\begin{IEEEproof}
This result is a direct consequence of \cite[VI.7.1, VI.7.2]{BhatiaBook}, where it is proved that $|\bT+\bR|\geq\prod_{j=1}^{N}\left(\lambda_{j}(T)+\lambda_{j}(R)\right)$, with $\{\lambda_{j}(T)\}_{j=1}^{N}$ and $\{\lambda_{j}(R)\}_{j=1}^{N}$ the eigenvalues of $\bT$ and $\bR$, respectively, arranged in decreasing order. As a consequence, since $|\bT+\bR|=|\bT||\bI+\bT^{-1}\bR|$, we have
\beq
\small
%\begin{split}
|\bT||\bI+\bT^{-1}\bR|\geq\prod_{j=1}^{N}\left(\lambda_{j}(T)+\lambda_{j}(R)\right)
=|\bT|\prod_{j=1}^{N}\left(1+\frac{\lambda_{j}(R)}{\lambda_{j}(T)}\right)
%\end{split}\;.
\notag
\eeq
\end{IEEEproof}

\begin{lemma}\label{Lemma:DiagonalX}
Consider the function $g:\bX\in {\cal H}^{n}\rightarrow g(\bX)\in \IR_{0}^{+}$, and assume that $g(\bGamma\bX\bGamma)=g(\bX)$ for any $n\times n$ diagonal matrix $\bGamma$ of the form
\beq\label{Eq:GammaMatrix}
\small
\bGamma={\rm diag}(\underbrace{1,\ldots,1}_{i-1},-1,\underbrace{1,\ldots,1}_{n-i})
\eeq
with $i\leq n$. Then, 
\begin{itemize}
\item if $g$ is concave, then it is maximized when $\bX$ is diagonal.
\item if $g$ is convex, then it is minimized when $\bX$ is diagonal.
\item if $g$ is linear, then, it remains unaltered if the eigenvector matrix of $\bX$ is set to the identity. 
\end{itemize}
\end{lemma}
\begin{IEEEproof}
This result is an extension of a technique first presented in \cite{hoesli2004monotonicity}. Assume $g$ is concave. Then, defining the matrix $\bX^{*}=\frac{1}{2}\bX+\frac{1}{2}\bGamma\bX\bGamma=\frac{1}{2}\bX+\frac{1}{2}\overline{\bX}$ we can write
\beq
\small
g(\bX^{*})=g\left(\frac{1}{2}\bX+\frac{1}{2}\overline{\bX}\right)\geq \frac{1}{2}g(\bX)+\frac{1}{2}g(\overline{\bX})=g(\bX)
\;.
\eeq
Now, $\bX^{*}$ has the same entries as $\bX$ except for the off-diagonal entries of the $i$-th row and column, that are set to zero. Therefore, iterating this procedure $n$ times leads to the conclusion that the maximizing $\bX$ must be diagonal. If $g$ is convex, the same procedure can be employed to show that 
\beq
\small
g(\bX^{*})=g\left(\frac{1}{2}\bX+\frac{1}{2}\overline{\bX}\right)\leq \frac{1}{2}g(\bX)+\frac{1}{2}g(\overline{\bX})=g(\bX)\;,
\eeq
thus proving that the minimizing $\bX$ must be diagonal. Finally, if $g$ is linear, combining the two previous cases we have $g(\bX^{*})\leq g(\bX)\leq g(\bX^{*})$, which implies $g(\bX^{*})=g(\bX)$. Iterating this procedure $n$ times yields $g(\bX)=g({\rm diag}(\bX))$, with ${\rm diag}(\bX)$ denoting the diagonal matrix with the same diagonal elements of $\bX$, which implies the thesis.
\end{IEEEproof}
\begin{lemma}\label{Lemma:ConcInv}
Let $\bM$ be a complex $N\times M$ matrix. Then, the matrix function $f(\bX)=(\bI_{N}+\bM\bX^{-1}\bM^{H})^{-1}$ is matrix-concave in $\bX$.
\end{lemma}
\begin{IEEEproof}
The thesis can be simply obtained by applying the matrix inversion lemma. 
\beq
\small
f(\bX)=\bI_{N}-\bM(\bX+\bM^{H}\bM)^{-1}\bM^{H}\;.
\eeq
The matrix function $(\bX+\bM^{H}\bM)^{-1}$ is matrix-convex, since it is a linear transformation of the matrix function $(\bX)^{-1}$, which is known to be matrix-convex \cite{marshall1979inequalities}. Then, the matrix-concavity of $f(\bX)$ immediately follows.
\end{IEEEproof}

\begin{lemma}\label{Lemma:ConcF}
For any $\nu\geq 0$, $M\times M$ diagonal, positive semidefinite matrices $\bL$ and $\bLambda$, the matrix-valued function $f(\bLambda)=\bLambda(\nu\bI_{M}+\bL\bLambda)^{-1}$ is matrix-concave.
\end{lemma}
\begin{IEEEproof}
First of all, denoting by $\blambda=\{\lambda_{i}\}_{i=1}^{M}$ and $\bell=\{\ell_{i}\}_{i=1}^{M}$ the diagonal vectors of $\bLambda$ and $\bL$, respectively, $f$ can be equivalently written as $f(\blambda)={\rm diag}\left(\frac{\lambda_{1}}{\nu+\ell_{1}\lambda_{1}},\ldots,\frac{\lambda_{M}}{\nu+\ell_{M}\lambda_{M}}\right)$. Then, exploiting the concavity of the function $\ds g_{m}(t)=\frac{t}{\nu+\ell_{m}t}$ for $t\geq 0$ and for all $m=1,\ldots,M$, 
we have 
\begin{gather}
\small
f(a\blambda_{1}+(1-a)\blambda_{2})\notag\\
\succeq a\;{\rm diag}\left(\frac{\lambda_{1,1}}{\nu+\ell_{1}\lambda_{1,1}},\ldots,\frac{\lambda_{1,M}}{\nu+\ell_{M}\lambda_{1,M}}\right)\notag\\
\small
+(1-a)\;{\rm diag}\left(\frac{\lambda_{2,1}}{\nu+\ell_{1}\lambda_{2,1}},\ldots,\frac{\lambda_{2,M}}{\nu+\ell_{M}\lambda_{2,M}}\right)\notag\\
\small
=af(\blambda_{1})+(1-a)f(\blambda_{2})\;,
\end{gather}
for all $a\in[0;1]$ and $\blambda_{1}=\{\lambda_{1,i}\}_{i=1}^{M}$, $\blambda_{2}=\{\lambda_{2,i}\}_{i=1}^{M}$ with non-negative components.
\end{IEEEproof}
\begin{lemma}\label{LemmaDerLogDet}
Let $\bM_{1}$ and $\bM_{2}$ be two $n\times n$ Hermitian matrices, and $x$ be a non-negative scalar. Then,
\begin{equation}
\small
\frac{d}{dx}{\rm log}|\bM_{1}+x\bM_{2}|={\rm tr}\left((\bM_{1}+x\bM_{2})^{-1}\bM_{2}\right)
\end{equation}
\end{lemma}
\begin{IEEEproof}
See \cite{jafar2004transmitter}.
\end{IEEEproof}

\bibliographystyle{IEEEtran}
%\nocite{*}
\bibliography{IEEEfull,Bibliography}

\begin{thebibliography}{10}
\providecommand{\url}[1]{#1}
\csname url@samestyle\endcsname
\providecommand{\newblock}{\relax}
\providecommand{\bibinfo}[2]{#2}
\providecommand{\BIBentrySTDinterwordspacing}{\spaceskip=0pt\relax}
\providecommand{\BIBentryALTinterwordstretchfactor}{4}
\providecommand{\BIBentryALTinterwordspacing}{\spaceskip=\fontdimen2\font plus
\BIBentryALTinterwordstretchfactor\fontdimen3\font minus
  \fontdimen4\font\relax}
\providecommand{\BIBforeignlanguage}[2]{{%
\expandafter\ifx\csname l@#1\endcsname\relax
\typeout{** WARNING: IEEEtran.bst: No hyphenation pattern has been}%
\typeout{** loaded for the language `#1'. Using the pattern for}%
\typeout{** the default language instead.}%
\else
\language=\csname l@#1\endcsname
\fi
#2}}
\providecommand{\BIBdecl}{\relax}
\BIBdecl

\bibitem{Cho2004}
J.~Cho and Z.~J. Haas, ``On the throughput enhancement of the downstream
  channel in cellular radio networks through multihop relaying,'' \emph{IEEE
  Journal on Selected Areas in Communications}, vol.~22, no.~7, pp. 1206--1219,
  September 2004.

\bibitem{Jiang2008}
P.~Jiang, J.~Bigham, and J.~Wu, ``Self-organizing relay stations in relay based
  cellular networks,'' \emph{Computer Communications}, vol.~31, no.~13, pp.
  2937--2945, August 2008.

\bibitem{Layer1Relay}
M.~Iwamura, H.~Takahashi, and S.~Nagata, ``Relay technology in
  {LTE}-advanced,'' \emph{NTT DOCOMO Technical Journal}, vol.~18, no.~2, pp.
  31--36, July 2010.

\bibitem{TseMIMO}
D.~Tse and P.~Viswanath, \emph{Fundamentals of Wireless Communications}.\hskip
  1em plus 0.5em minus 0.4em\relax Cambridge University Press, 2005.

\bibitem{munoz2007linear}
O.~Munoz-Medina, J.~Vidal, and A.~Agust{\'\i}n, ``Linear transceiver design in
  nonregenerative relays with channel state information,'' \emph{IEEE
  Transactions on Signal Processing}, vol.~55, no.~6, pp. 2593--2604, June
  2007.

\bibitem{chae2008mimo}
C.~Chae, T.~Tang, R.~Heath, and S.~Cho, ``{MIMO} relaying with linear
  processing for multiuser transmission in fixed relay networks,'' \emph{IEEE
  Transactions on Signal Processing}, vol.~56, no.~2, pp. 727--738, February
  2008.

\bibitem{Vaze2011}
R.~Vaze and R.~W. Heath, ``On the capacity and diversity multiplexing tradeoff
  of the two-way relay channel,'' \emph{IEEE Transactions on Information
  Theory}, vol.~57, no.~7, pp. 4219--4234, July 2011.

\bibitem{Rong2009}
Y.~Rong and Y.~Hua, ``Optimality of diagonalization of multihop {MIMO}
  relays,'' \emph{IEEE Transactions on Wireless Communications}, vol.~8,
  no.~12, pp. 6068--6077, December 2009.

\bibitem{Calvo2009}
E.~Calvo, J.~Vidal, and J.~R. Fonollosa, ``Optimal resource allocation in
  relay-assisted cellular networks with partial {CSI},'' \emph{IEEE
  Transactions on Signal Processing}, vol.~57, no.~7, pp. 2809--2823, July
  2009.

\bibitem{yu-power}
Y.~Yu and Y.~Hua, ``Power allocation for a {MIMO} relay system with
  multiple-antenna users,'' \emph{IEEE Transactions on Signal Processing},
  vol.~58, no.~5, pp. 2823--2835, April 2010.

\bibitem{MIMOAFRelayPowerA}
I.~Hammerstrom and A.~Wittneben, ``{Power allocation schemes for
  amplify-and-forward MIMO-OFDM relay links},'' \emph{IEEE Transactions on
  Wireless Communications}, vol.~6, no.~8, pp. 2798--2802, August 2007.

\bibitem{WangTao2012}
R.~Wang and M.~Tao, ``Joint source and relay precoding designs for {MIMO}
  two-way relaying based on {MSE} criterion,'' \emph{IEEE Transactions on
  Signal Processing}, vol.~60, no.~3, pp. 1352--1365, March 2012.

\bibitem{Heath2013}
K.~Truong, P.~Sartori, and R.~Heath, ``Cooperative algorithms for {MIMO}
  amplify-and-forward relay networks,'' \emph{IEEE Transactions on Signal
  Processing}, vol.~61, no.~5, pp. 1272--1287, March 2013.

\bibitem{ZapTWC12}
A.~Zappone and E.~Jorswieck, ``{Resource Allocation in Amplify-and-Forward
  Relay-Assisted DS/CDMA Systems},'' \emph{IEEE Transactions on Wireless
  Communications}, vol.~11, no.~4, pp. 1271--1276, April 2012.

\bibitem{commmag_si}
``Special issue on energy efficiency in communications,'' \emph{Communications
  Magazine, IEEE}, vol.~48, no.~11, November 2010.

\bibitem{jsac_si}
``Special issue on energy-efficient wireless communications,'' \emph{IEEE
  Journal on Selected Areas in Communications}, vol.~29, no.~8, September 2011.

\bibitem{Pickavet}
M.~Pickavet, W.~Vereecken, S.~Demeyer, P.~Audenaert, B.~Vermeulen, C.~Develder,
  D.~Colle, B.~Dhoedt, and P.~Demeester, ``Worldwide energy needs for ict: The
  rise of power-aware networking,'' in \emph{Advanced Networks and
  Telecommunication Systems, 2008. ANTS '08. 2nd International Symposium on},
  December 2008, pp. 1 --3.

\bibitem{MultihopMIMORelayRong11}
Y.~Rong, ``{Multihop nonregenerative MIMO relays-QoS considerations},''
  \emph{IEEE Transactions on Signal Processing}, vol.~59, no.~1, pp. 290--303,
  January 2011.

\bibitem{PowerallMIMOAF2012}
L.~Sanguinetti and A.~A. D'Amico, ``{Power allocation in two-hop
  amplify-and-forward MIMO relay systems with QoS requirements},'' \emph{IEEE
  Transactions on Signal Processing}, vol.~60, no.~5, pp. 2494--2507, May 2012.

\bibitem{ZapOFDMA}
S.~Buzzi, G.~Colavolpe, D.~Saturnino, and A.~Zappone, ``Potential games for
  energy-efficient power control and subcarrier allocation in uplink multicell
  ofdma systems,'' \emph{Selected Topics in Signal Processing, IEEE Journal
  of}, vol.~6, no.~2, pp. 89 --103, April 2012.

\bibitem{WidelyLinear}
S.~Buzzi, H.~Poor, and A.~Zappone, ``Transmitter {W}aveform and
  {W}idely-{L}inear {R}eceiver {D}esign: {N}on-cooperative {G}ames for
  {W}ireless {M}ultiple-{A}ccess {N}etworks,'' \emph{IEEE Transactions on
  Information Theory}, vol.~56, no.~10, pp. 4874--4892, October 2010.

\bibitem{Betz2008}
S.~Betz and H.~V. Poor, ``{Energy Efficient Communications in {CDMA} Networks:
  A Game Theoretic Analysis Considering Operating Costs},'' \emph{IEEE
  Transactions on Signal Processing}, vol.~56, no.~10, pp. 5181--5190, October
  2008.

\bibitem{Meshkati2007}
F.~Meshkati, S.~C. Schwartz, and H.~V. Poor, ``{Energy-Efficient Resource
  Allocation in Wireless Networks},'' \emph{IEEE Signal Processing Magazine},
  vol.~24, no.~3, pp. 58--68, May 2007.

\bibitem{Miao2011}
G.~Miao, N.~Himayat, G.~Y. Li, and S.~Talwar, ``{Distributed Interference-Aware
  Energy-Efficient Power Optimization},'' \emph{IEEE Transactions on Wireless
  Communications}, vol.~10, no.~4, pp. 1323--1333, April 2011.

\bibitem{Isheden2011}
C.~Isheden, Z.~Chong, E.~Jorswieck, and G.~Fettweis, ``{Framework for
  Link-Level Energy Efficiency Optimization with Informed Transmitter},''
  \emph{IEEE Transactions on Wireless Communications}, vol.~11, no.~8, pp.
  2946--2957, August 2012.

\bibitem{BuzziMIMO}
S.~Buzzi, H.~V. Poor, and D.~Saturnino, ``Energy-efficient resource allocation
  in multiuser {MIMO} systems: a game-theoretic framework,'' in \emph{Proc. of
  the 16th European Signal Processing Conference (EUSIPCO 2008), Switzerland},
  August 2008.

\bibitem{belmega2011energy}
E.~Belmega and S.~Lasaulce, ``Energy-efficient precoding for multiple-antenna
  terminals,'' \emph{IEEE Transactions on Signal Processing}, vol.~59, no.~1,
  pp. 329--340, 2011.

\bibitem{MIMOBroadcastEE}
J.~Xu and L.~Qiu, ``Energy efficiency optimization for {MIMO} broadcast
  channels,'' \emph{IEEE Transactions on Wireless Communications}, vol.~12,
  no.~2, pp. 690--701, February 2013.

\bibitem{Zap2011Letter}
A.~Zappone, S.~Buzzi, and E.~Jorswieck, ``{E}nergy-{E}fficient power control
  and receiver design in relay-assisted {DS/CDMA} wireless networks via game
  theory,'' \emph{IEEE Communications Letters}, vol.~15, no.~7, pp. 701--703,
  July 2011.

\bibitem{ZapponeTWC}
A.~Zappone, Z.~Chong, E.~Jorswieck, and S.~Buzzi, ``Energy-aware competitive
  power control in relay-assisted interference wireless networks,'' \emph{IEEE
  Transactions on Wireless Communications}, vol.~12, no.~4, pp. 1860--1871,
  April 2013.

\bibitem{soysal2007optimum}
A.~Soysal and S.~Ulukus, ``Optimum power allocation for single-user {MIMO} and
  multi-user {MIMO-MAC} with partial {CSI},'' \emph{IEEE Journal on Selected
  Areas in Communications}, vol.~25, no.~7, pp. 1402--1412, September 2007.

\bibitem{soysal2009optimality}
------, ``Optimality of beamforming in fading {MIMO} multiple access
  channels,'' \emph{IEEE Transactions on Communications}, vol.~57, no.~4, pp.
  1171--1183, April 2009.

\bibitem{ZapRelayMIMO2013}
A.~Zappone and E.~A. Jorswieck, ``Resource allocation in relay-assisted {MIMO}
  {MAC} systems with statistical {CSI},'' \emph{Physical Communications}, 2013,
  in press.

\bibitem{TaoImperfect2012}
J.~Zou, H.~Luo, M.~Tao, and R.~Wang, ``Joint source and relay optimization for
  non-regenerative {MIMO} two-way relay systems with imperfect {CSI},''
  \emph{IEEE Transactions on Wireless Communications}, vol.~11, no.~9, pp.
  3305--3315, September 2012.

\bibitem{Chong2011c}
Z.~Chong and E.~A. Jorswieck, ``Energy-efficient power control for {MIMO}
  time-varying channels,'' in \emph{IEEE Online Green Communications
  Conference}, September 2011.

\bibitem{AlfanoChong}
G.~Alfano, Z.~Chong, and E.~A. Jorswieck, ``Energy-efficient power control for
  {MIMO} channels with partial and full {CSI},'' in \emph{2012 International
  ITG Workshop on Smart Antennas (WSA)}, March 2012.

\bibitem{CaoCAMAD12}
P.~Cao, Z.~Chong, Z.~K.~M. Ho, and E.~Jorswieck, ``{Energy-Efficient Power
  Allocation for Amplify-and-Forward MIMO Relay Channel},'' in \emph{Proc. of
  IEEE CAMAD}, September 2012.

\bibitem{EEtwowayRelay}
C.~Sun and C.~Yang, ``Energy efficiency analysis of one-way and two-way relay
  systems,'' \emph{EURASIP Journal on Wireless Communications and Networking},
  vol. February, pp. 1--18, 2012.

\bibitem{RFPc1Cui2005}
S.~Cui, A.~Goldsmith, and A.~Bahai, ``{Energy-constrained modulation
  optimization},'' \emph{IEEE Transactions on Wireless Communications}, vol.~4,
  no.~5, pp. 2349---2360, September 2005.

\bibitem{BertsekasNonLinear}
D.~Bertsekas, \emph{Nonlinear Programming}.\hskip 1em plus 0.5em minus
  0.4em\relax Athena Scientific, 1999.

\bibitem{FracProgSS1983}
S.~Schaible, ``{Fractional programming},'' \emph{Zeitschrift f{\"u}r Operations
  Theory and Applications}, vol.~27, no.~1, pp. 347--352, 1983.

\bibitem{NonlinearFracProg}
W.~Dinkelbach, ``{On nonlinear fractional programming},'' \emph{Management
  Science}, vol.~13, no.~7, pp. 492--498, March 1967.

\bibitem{shiu2000fading}
D.~Shiu, G.~Foschini, M.~Gans, and J.~Kahn, ``{Fading correlation and its
  effect on the capacity of multielement antenna systems},'' \emph{IEEE
  Transactions on Communications}, vol.~48, no.~3, pp. 502--513, March 2000.

\bibitem{jafar2004transmitter}
S.~Jafar and A.~Goldsmith, ``Transmitter optimization and optimality of
  beamforming for multiple antenna systems,'' \emph{IEEE Transactions on
  Wireless Communications}, vol.~3, no.~4, pp. 1165--1175, July 2004.

\bibitem{jorswieck2004channel}
E.~Jorswieck and H.~Boche, ``Channel capacity and capacity-range of beamforming
  in {MIMO} wireless systems under correlated fading with covariance
  feedback,'' \emph{IEEE Transactions on Wireless Communications}, vol.~3,
  no.~5, pp. 1543--1553, September 2004.

\bibitem{LuoWMMSE}
Q.~Shi, M.~Razaviyayn, Z.~Q. Luo, and C.~He, ``{An Iteratively Weighted MMSE
  Approach to Distributed Sum-Utility Maximization for a MIMO Interfering
  Broadcast Channel},'' \emph{IEEE Transactions on Signal Processing}, vol.~59,
  no.~9, pp. 4331--4340, September 2011.

\bibitem{DharmawansaBeam2011}
P.~Dharmawansa, M.~R. McKay, R.~K. Mallik, and K.~B. Letaief, ``{Ergodic
  Capacity and Beamforming Optimality for Multi-Antenna Relaying with
  Statistical CSI},'' \emph{IEEE Transactions on Communications}, vol.~59,
  no.~8, pp. 2119--2131, August 2011.

\bibitem{AhamadiBeam2011}
S.~Al-Ahmadi and H.~Yanikomeroglu, ``On the beamforming optimality range in
  {TIMO} channels with common and individual input power constraints,''
  \emph{IEEE Transactions on Communications}, vol.~59, no.~3, pp. 648--651,
  March 2011.

\bibitem{BeamMISO}
E.~Vagenas, G.~S. Paschos, and S.~A. Kotsopoulo, ``Beamforming capacity
  optimization for {MISO} systems with both mean and covariance feedback,''
  \emph{IEEE Transactions on Wireless Communications}, vol.~11, no.~9, pp.
  2994--3001, September 2011.

\bibitem{marshall1979inequalities}
A.~Marshall and I.~Olkin, \emph{Inequalities: theory of majorization and its
  applications}.\hskip 1em plus 0.5em minus 0.4em\relax Academic Press New
  York, 1979.

\bibitem{BhatiaBook}
R.~Bhatia, \emph{Matrix Analysis}.\hskip 1em plus 0.5em minus 0.4em\relax
  Springer, 1997.

\bibitem{hoesli2004monotonicity}
D.~Hoesli, Y.~Kim, and A.~Lapidoth, ``Monotonicity results for coherent {MIMO}
  {R}ician channels,'' \emph{IEEE Transactions on Information Theory}, vol.~51,
  no.~12, pp. 4334--4339, December 2005.

\end{thebibliography}

\begin{figure}[htbp]
\centering
\includegraphics[scale=0.45]{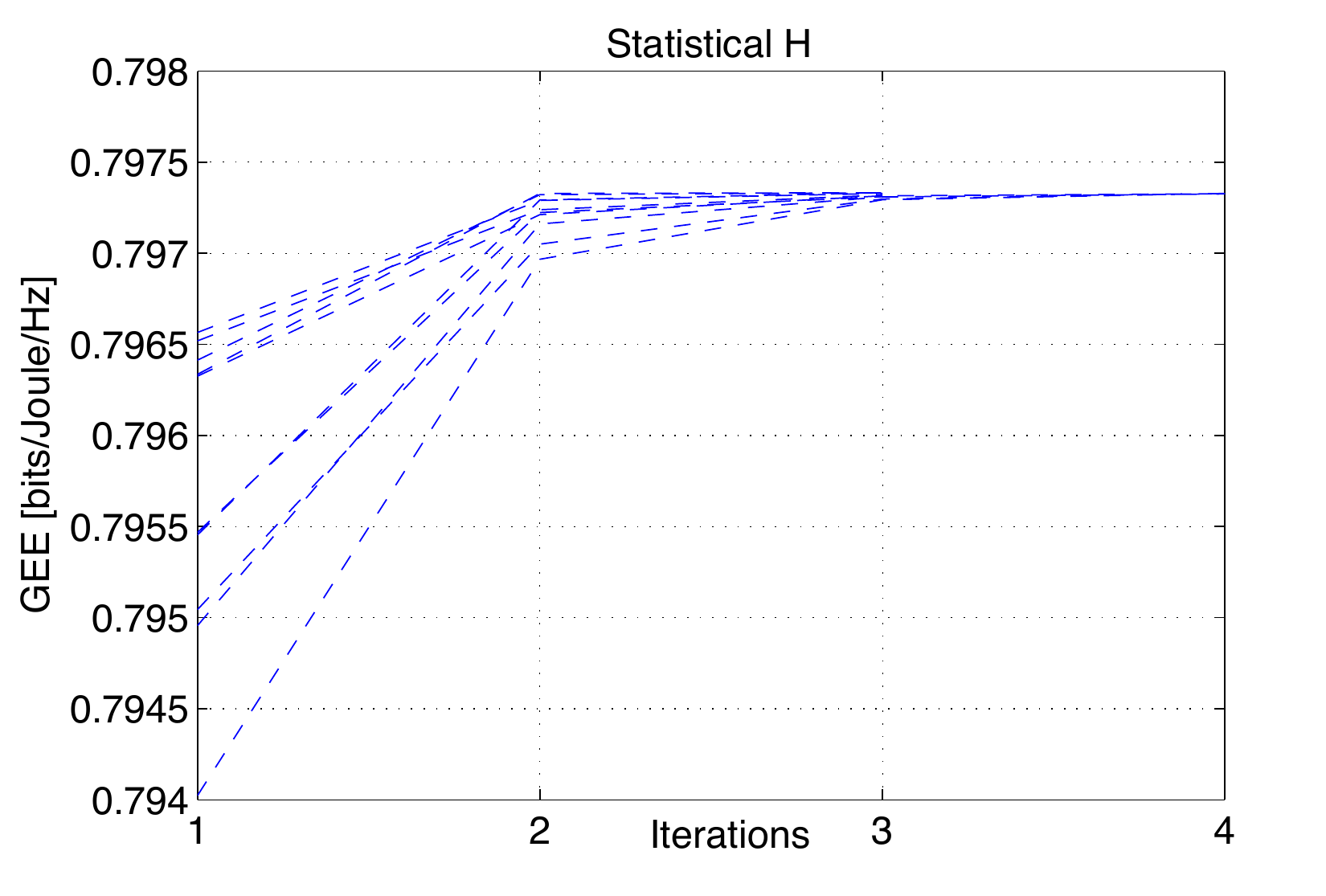}
\caption{Number of iterations for the alternating maximization algorithm to converge with statistical CSI on $\bH$. The algorithm has been started from $10$ different initialization points.} \label{fig:1}
\end{figure}

\begin{figure}[htbp]
\centering
\includegraphics[scale=0.45]{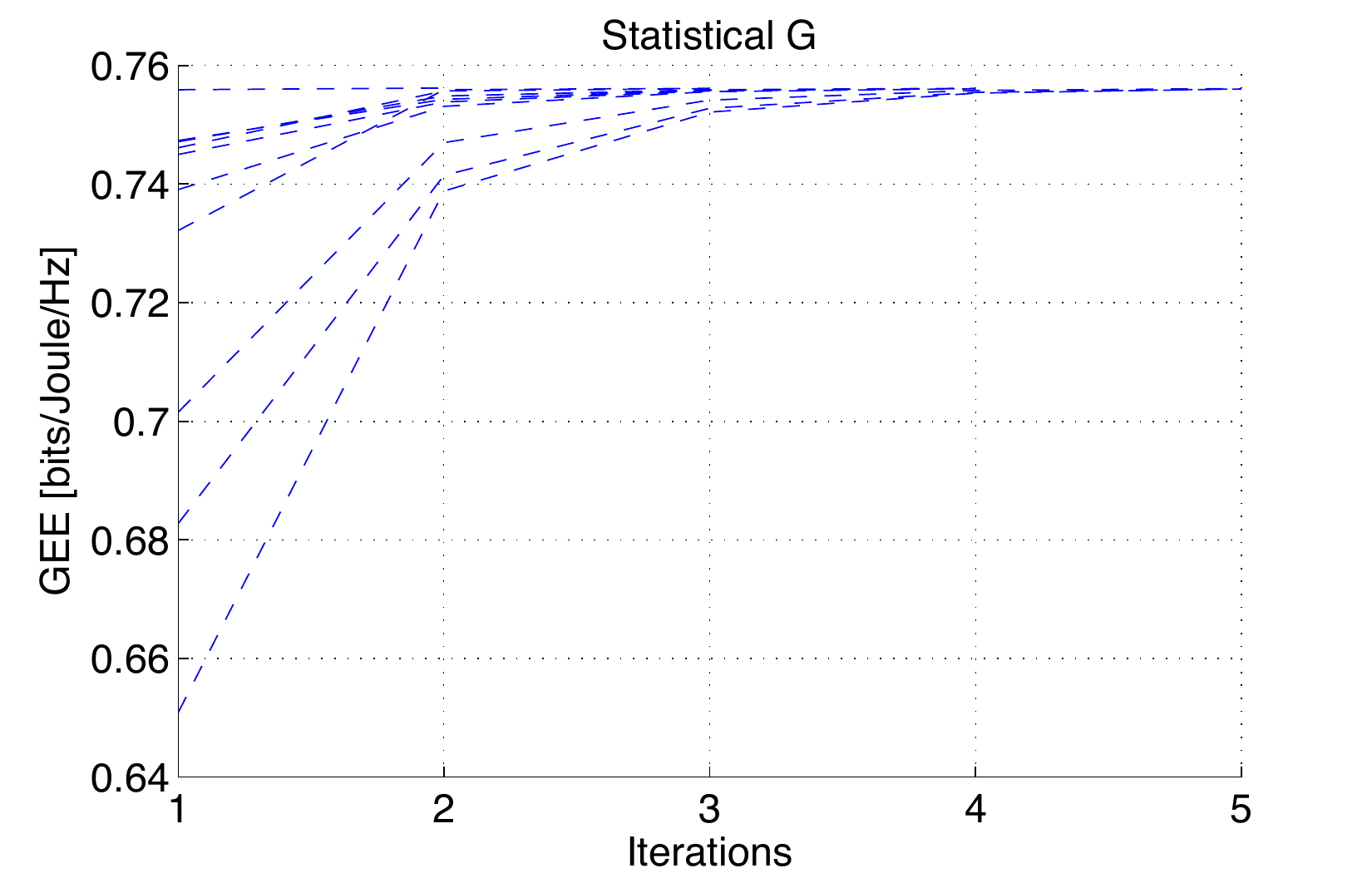}
\caption{Number of iterations for the alternating maximization algorithm to converge with statistical CSI on $\bG$. The algorithm has been started from $10$ different initialization points.} \label{fig:2}
\end{figure}

\begin{figure}[htbp]
\centering
\includegraphics[scale=0.5]{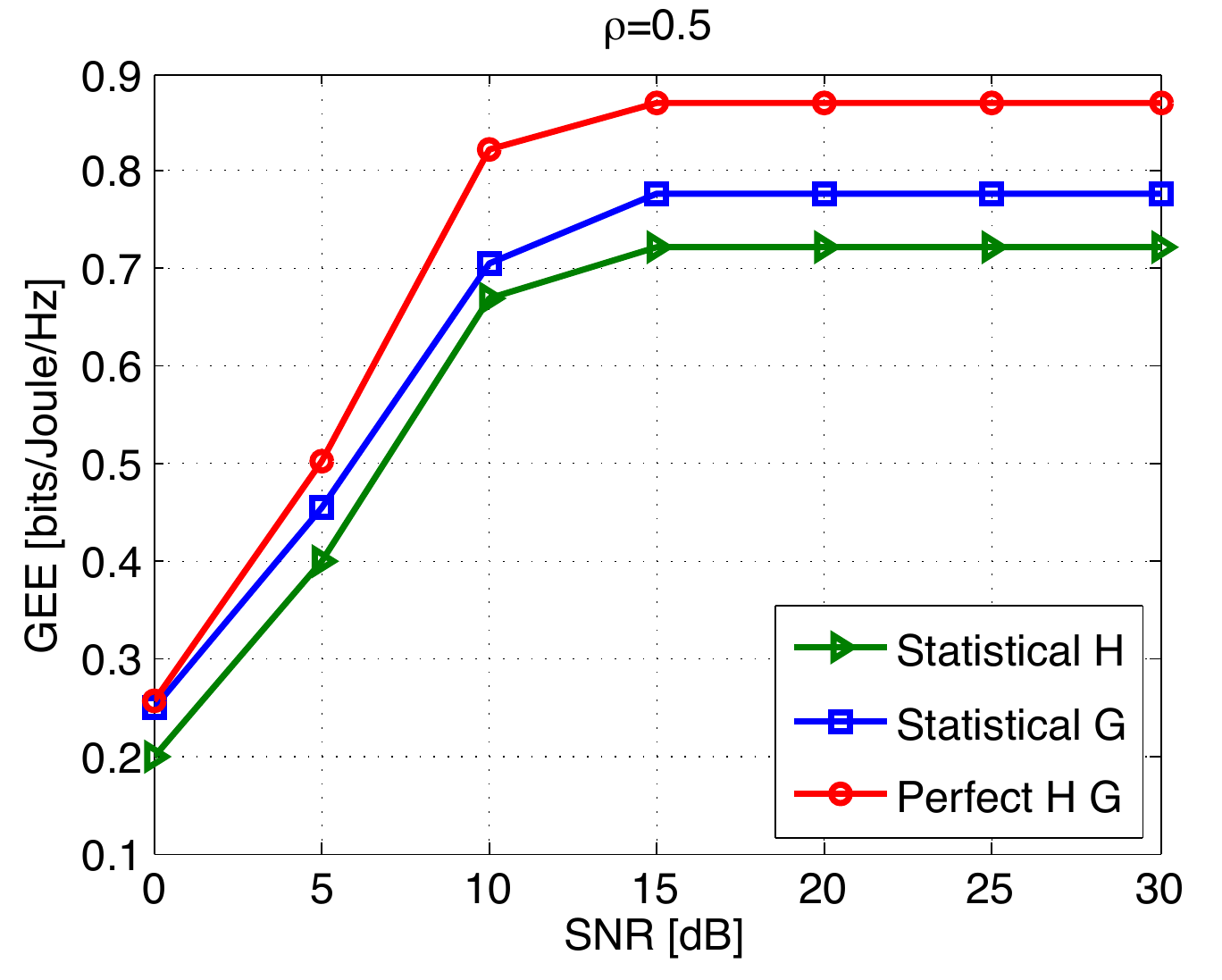}
\caption{$\rho=0.5$. Achieved average GEE for: a) GEE maximization with perfect CSI; b) GEE maximization with statistical CSI on $\bH$; c) GEE maximization with statistical CSI on $\bG$.} \label{fig:3}
\end{figure}

\begin{figure}[htbp]
\centering
\includegraphics[scale=0.5]{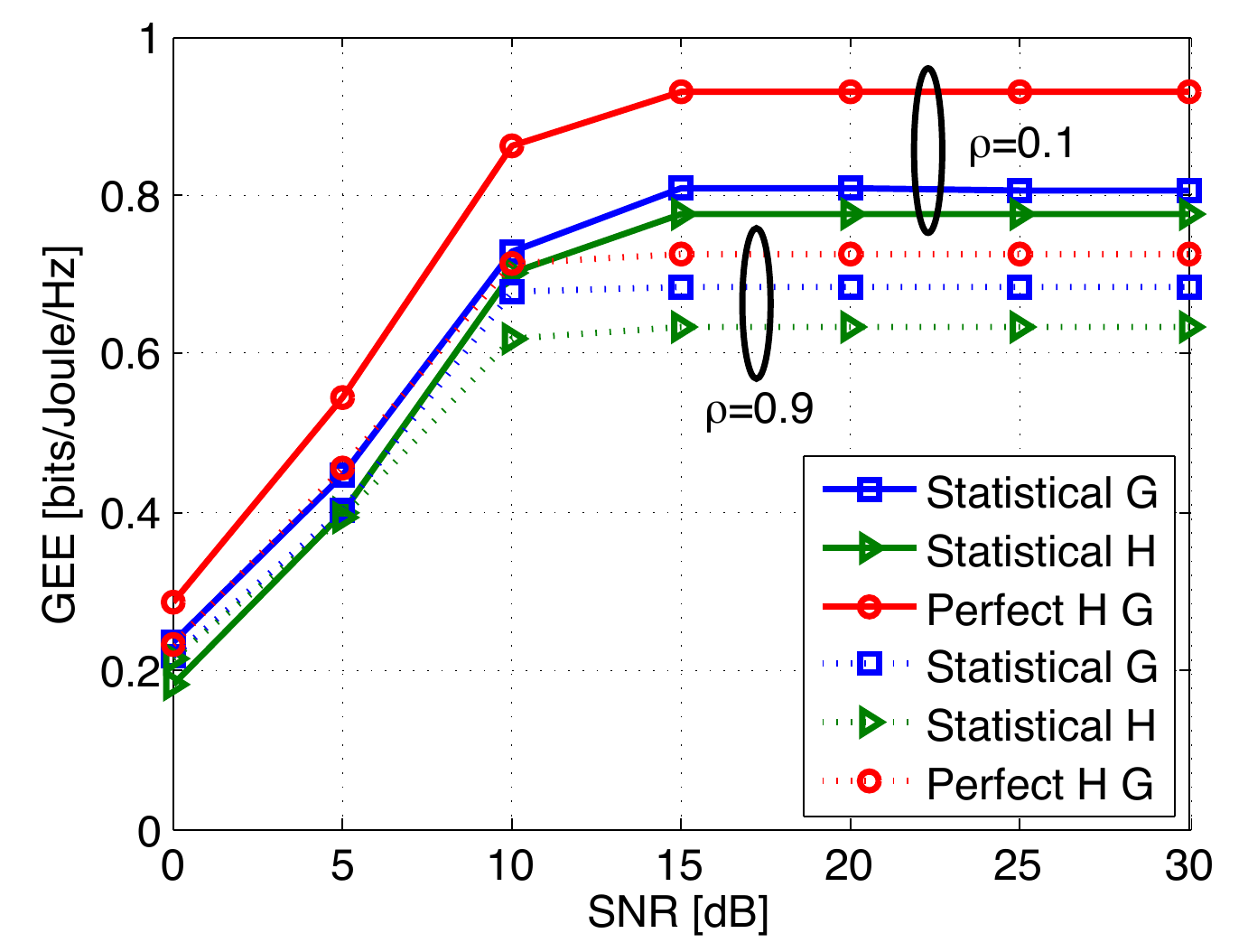}
\caption{$\rho=0.1;0.9$. Achieved average GEE for: a) GEE maximization with perfect CSI; b) GEE maximization with statistical CSI on $\bH$; c) GEE maximization with statistical CSI on $\bG$.} \label{fig:4}
\end{figure}

\begin{figure}[htbp]
\centering
\includegraphics[scale=0.37]{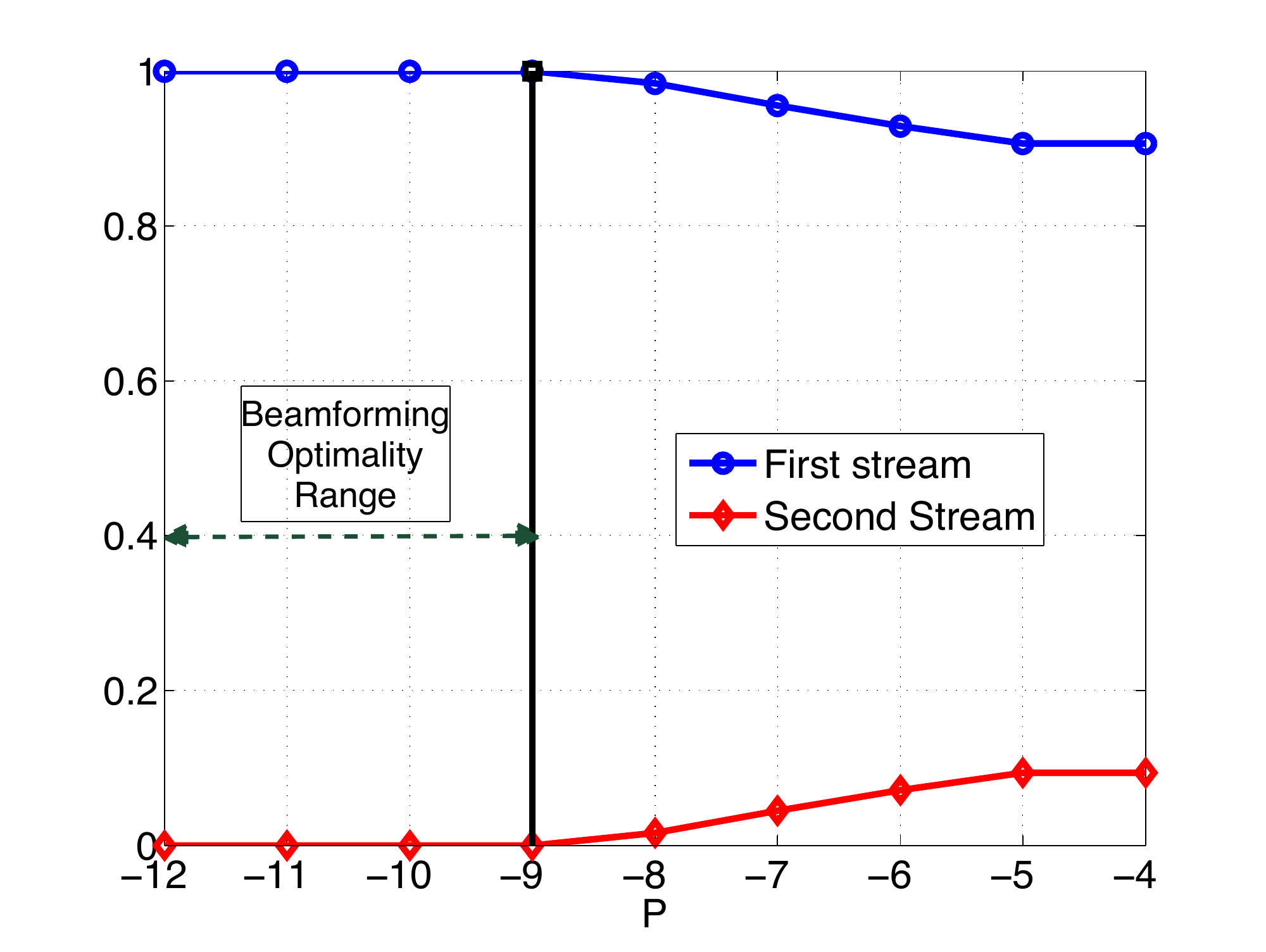}
\caption{FP beamforming optimality range for statistical CSI on $\bH$. For $P\leq -9$dBW, FP beamforming is optimal.} \label{fig:5}
\end{figure}

\end{document}